\documentclass{article}
\usepackage[utf8]{inputenc}
\usepackage[margin=2cm]{geometry}
\usepackage{authblk}
\usepackage{setspace}
\usepackage{graphicx}
\graphicspath{{./figures/}}
\usepackage{subcaption}
\usepackage{amsmath}
\usepackage{booktabs}
\usepackage{hyperref}
\usepackage{todonotes}
\usepackage{pdfcomment}
\usepackage[switch]{lineno}       
\usepackage[
  style=nejm, 
  citestyle=numeric-comp,
  sorting=none
]{biblatex}
\newcommand{\comment}[1]{}

\newcommand{\fref}[1]{Fig. \ref{#1}}
\newcommand{\Fref}[1]{Figure \ref{#1}}

\title{ A Framework for Intrinsic Poincaré Sections and Phase-Space Manifold Visualization: A Case Study of the Planar Elastic Pendulum}

\author[1]{Rafael Salandin Moraes}
\author[1*]{Florian Steffen Günther}

\affil[1]{Department of Physics, UNESP, Rio Claro, Brazil.}
\affil[*]{Address correspondence to: florian.gunther@unesp.br}
\date{\today}

\begin{document}
\begin{@twocolumnfalse}
\maketitle
\begin{abstract}
	The global phase-space organization of non-linear Hamiltonian systems is traditionally visualized using Poincaré sections. However, rigid choices of sectioning hyperplanes often introduce geometric distortions and coordinate artifacts that obscure or clip fundamental invariant structures. Here, we present a multi-mapping analysis of the planar elastic pendulum to systematically overcome these visual and structural limitations. We implement a comparative framework utilizing inverted phase-space mappings that resolve the dense packing of invariant curves near chaotic boundaries, uncovering an apparent separatrix trajectory hidden in standard views. Leveraging the system's vertical symmetry axis, we derive two novel classes of customized canonical transformations that align the sectioning condition with the underlying force field and invariant trajectories, respectively. We demonstrate that the force-line section balances phase-space density representation near equilibrium and exposes a curvature-driven, hexagon-like boundary deformation. Concurrently, the trajectory-aligned section unrolls highly curved invariant manifolds into a regular grid. When combined with inverted mapping, this trajectory-based approach acts as a structural coordinate zoom, minimizing local metric distortions and shifting delicate, higher-order satellite islands directly into the focal center. Our results demonstrate that relying on a single slice is insufficient to capture complex non-linear dynamics; instead, utilizing at least two orthogonal, field-conforming sections provides a superior, distortion-free diagnostic tool for characterizing structural stability, resonance chains, and global transport barriers in multi-degree-of-freedom systems.
\end{abstract}
\vspace{1em}
\end{@twocolumnfalse}


\section{Introduction}
One of the greatest challenges in the analysis of nonlinear systems is the co-existence and intricate division of the phase space into regular and chaotic trajectories. 
For a given set of control parameters, these different dynamical regimes often intermingle in complex, highly fragmented and often fractal patterns. 
To visualize and understand this global phase-space organization,  Poincaré sections are the standard tool in nonlinear science.
By recording the coordinates each time a trajectory crosses a specific hyperplane, this method reduces the dimensionality and provides a visual map of the othervise not visuabilisable phase space.
The specific mapping, that is thereby considered, however, strongly depends on the choice of the section, and may or may not be incapable or inefficient to present all the dynamics of interest.
It is therefore meaningful to carefully considered how to choose the section and/or simultaneously consider and compare different sections.

\begin{figure}[!htb]
	\centering
	\includegraphics[width=0.5\linewidth]{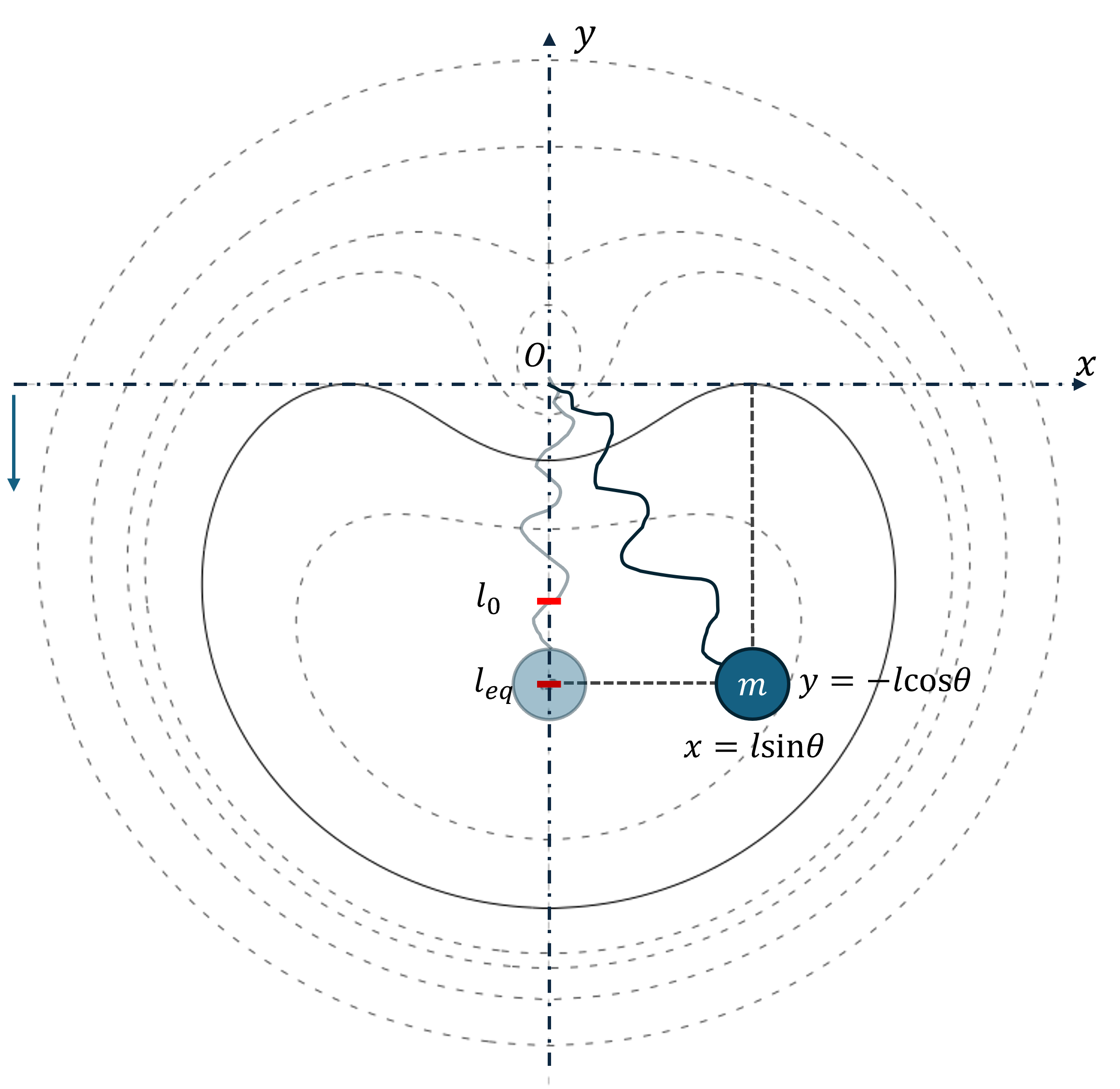}
	\caption{Elastic pendulum schematic detailing the system's equipotential curves (dashed gray lines) and the specific equipotential curve analyzed in this work (solid line). The geometric mapping between Cartesian and polar coordinates is also shown.}
	\label{fig:placeholder}
\end{figure}

In this work, a detailed analysis how the choices of sections impact the properties visible in the corresponding mappings is conducted. 
For this,  the planar elastic pendulum  is considered which is a simple Hamiltonian system with two degrees of freedom, consisting of a mass connected to a pivot by an ideal spring, see \Fref{fig:placeholder}.
This system serves as an ideal system for our study, due to the following reasons:
First, the phyisical interpretation of the model is a rather simple hamiltonian system. 
It arrises from  the simple harmonic osilator  of the mass-spring system when lateral movment is not neglegted.
Second, despite its rather simple physical description, the motion for given initial conitions is goverend by the coupling between the pendulum and the spring oscilation, the interplay of which causes a remarkably rich and complex dynamics, including autoparametric resonance.\cite{vitt1933,NUNEZYEPEZ1990101, vanderWeele1996, Tarigo2026}
Third, the hamiltonian arising from the spring-mass system can also be obtained as limit for other pysical system under certain approximation.\cite{tselman1970, NUNEZYEPEZ1990101,carretero1994 ,CDESOUSA20181110}
Last and most interesting, the system of the spring pendulum serves as an ideal example showcasing that the coice of the best section is not well defined or obvious, since different works reported over the past years partially consider differnt sections.

The elastic pendulum has a long history in physics, first appearing in the literature in a 1933 paper by Vitt and Gorelik \cite{vitt1933}, where it served as an analogy for vibrations in CO$_2$ molecules.
This and other early works, however, focused only on describing stable, quasi-periodic trajectories through analytical approximations.
It was not until the 1990s that modern computing allowed researchers to study chaotic trajectories.
 In their work, H.~N. N{\'u}{\~n}ez-Y{\'e}pez and co-workers \cite{NUNEZYEPEZ1990101} presented the first study showing the transition between regular and chaotic motion using Poincar{\'e} maps. 
 Due to their choice of coordinates to set up the equations of motion, their map used a horizontal section in the configuration space that passes through the equilibrium position. 
Shortly after, Cuerno \textit{et al.} \cite{Cuerno1992} presented a study on the deterministic chaos of the elastic pendulum. 
They visualized the chaotic sea and KAM stability islands presenting Poincar{\'e} mapping with differnt  properties as that of H.~N. N{\'u}{\~n}ez-Y{\'e}pez \textit{et al.}. 
Without explicitly mentioning it, their section differed from that of H.~N. N{\'u}{\~n}ez-Y{\'e}pez because they considered a vertical section passing through the pivot point. 
Utilizing this same vertical section, van der Weele and de Kleine \cite{vanderWeele1996}, as well as Kuznetsov \cite{Kuznetsov1999}, investigated the order-chaos-order transition. 
Unlike Cuerno et al., these studies strongly focused on high-energy trajectories with rotational motion around the pivot point of the spring.
In addition to the vertical section, Kuznetsov \cite{Kuznetsov1999} also presented Poincar{\'e} sections through the phase space where  momenta rather than a coordinate was fixed.
In more recent publication by  Anurag et al. \cite{Anurag2020}, the order-chaos-order transition was studied using a forulation of the problem in polar coordinates. 
This resulted in the consideration of a circular section of the configuration space.
In another study, the same authors analyzed the elastic pendulum in 3D also discussing the planar version, in which a horizontal section through the pivot point \cite{anurag2022} was used to generate poincaré maps.

Over the past years, several studies on the elastic pendulum have been published typically considering either of these different Poincar{\'e} sections. 
In general, a clear focus was given on the vertical \cite{NUNEZYEPEZ1990101, vanderWeele1996, Kuznetsov1999, Anurag2020, Tarigo2026} and the horizontal sections of the configuration space \cite{CDESOUSA20181110,CDESOUSA2022128481}.
For an overall characterization of nonlinear dynamics---such as determining whether a trajectory is regular or chaotic---the specific choice of the section is often irrelevant.
However, it is crucial to have a fundamental understanding of which specific instances in the configuration space are reflected in the corresponding map.
This understanding is particularly important when trying to connect and compare the reported properties across different published works.
The aim of the present work is therfore to compare these individual mappings and correlate their visual characteristics with the geometric meaning of each specific section choice.
Based on these observations, we suggest new types of sections that build upon the properties of traditional ones.
To formulate these new sections using the standard condition where a single coordinate is constant, we introduce new coordinate systems tailored to intrinsically describe these novel sections.

The remainder of the work is organized as follows. In section 2, we start by revising the mathematical formulation of the elastic pendulum in both Carthesian and Polar formulation and introduce proper dimensionless coordinates and deduce which Poincaré section in configuration space follow from naiv consideration of constante coordinates. 
In section 3, after summerizing our computational approach, we classify the difderent types of trajectories which occure for the parameter combination $(\mathcal H,\omega)$ considered throught this study.
This classification is crusial as it serves as guide to follow, how certain phasespace structure appear in different sections.
In section 4, we revise and compare the Poincaré mapping obtained from the traditional section and identify which features are or are not presented by the individial section. In doing so, the concept of inverted section is introduced, which mapps the same section, however considered the opposite momentum,
Finally, in Section 6, we formulate new sections conditions which arise from the intrinsic properties of the vertical section. 
For both proper canonical transformation are formulated in order to present these section again as naiv section, considering one coordinate constant.
Concluding the work, the major finidings are summarized in the final section 7 together with motivations for follow up research outlooks.
  
\section{Traditional Coordinates and Associated Poincaré sections}
\subsection{Cartesian Formulation}
Using two-dimensional Cartesian coordinates $(x, y)$ with the origin fixed at the pivot point is a straightforward choice for investigating the dynamics of the elastic pendulum. In this case, the required expressions for potential and kinetic energy are directly at hand, providing the expression for the Hamilton function as:
\begin{equation}
	H(x, y, p_x, p_y) = \frac{p_x^2 + p_y^2}{2m} + \frac{1}{2}k \left( \sqrt{x^2 + y^2} - l_0 \right)^2 + mgy
\end{equation}
Here, $m$ is the mass, $k$ denotes the spring constant, $l_0$ is the unstretched natural length of the spring, and $g$ is the acceleration due to gravity pointing in negative $y$-direction.

To reduce the number of independent parameters and to generalize the numeric description of the dynamics, it is meaningful to transform the system into dimensionless variables. Following the convention of choosing the natural length $l_0$ as the characteristic length scale and the characteristic frequency of the spring oscillation $\omega_S = \sqrt{k/m}$ to scale the time $t$, the following dimensionless quantities are defined:
\begin{equation} \label{eq:coord_cart}
	q_{\textnormal{c}1} = \frac{x}{l_0}, \quad q_{\textnormal{c}2} = \frac{y}{l_0} , \quad
	p_{\textnormal{c}1} = \frac{p_x}{m l_0 \omega_S}, \quad p_{\textnormal{c}2} = \frac{p_y}{m l_0 \omega_S}, \quad \tau = t \omega_S = t \sqrt{\frac{k}{m}} \\
\end{equation}
Applying this scaling to the Hamiltonian and dividing by the energy unit $E_{\rm ref} = k l_0^2$, we obtain the dimensionless Hamiltonian $\mathcal{H} = H / (k l_0^2)$:
\begin{equation}
	\mathcal{H} = \frac{1}{2}(p_{\textnormal{c}1}^2 + p_{\textnormal{c}2}^2) + \frac{1}{2} \left( \sqrt{q_{\textnormal{c}1}^2 + q_{\textnormal{c}2}^2} - 1 \right)^2 + \omega^2 q_{\textnormal{c}2}
\end{equation}
where the dimensionless control parameter $\omega^2$ is defined as
\begin{equation}
	\omega^2 = \frac{mg}{k l_0} = \frac{g/l_0}{k/m} = \frac{\omega_P^2}{\omega_S^2} \, .
\end{equation}
This definition follows that of Acosta-Zamora {\it et al.} \cite{AcostaZamora2024}, but differs from other works \cite{CDESOUSA20181110, Anurag2020, Tarigo2026} in which the reference length for the pendulum oscillation is with respect to the stretched equilibrium length of the spring under load.

The temporal evolution of the system is given by the respective canonical equations:
\begin{equation}
	\left\{
	\begin{aligned}
		\dot{q}_{\textnormal{c}1} &= p_{\textnormal{c}1}, \\
		\dot{q}_{\textnormal{c}2} &= p_{\textnormal{c}2}, \\
		\dot{p}_{\textnormal{c}1} &=
		\left(1- \sqrt{q_{\textnormal{c}1}^2 + q_{\textnormal{c}2}^2}  \right)
		\frac{q_{\textnormal{c}1}}
		{\sqrt{q_{\textnormal{c}1}^2 + q_{\textnormal{c}2}^2}}, \\
		\dot{p}_{\textnormal{c}2} &=
		\left(1- \sqrt{q_{\textnormal{c}1}^2 + q_{\textnormal{c}2}^2} \right)
		\frac{q_{\textnormal{c}2}}
		{\sqrt{q_{\textnormal{c}1}^2 + q_{\textnormal{c}2}^2}}
		-\omega^2 .
	\end{aligned}
	\right.
	\label{Eq:motionCart}
\end{equation}

\subsection{Polar Formulation}
Alternatively to the use of Cartesian coordinates, the system is equally straightforward to formulate in polar coordinates:
\begin{equation}
	H(l, \vartheta, p_l, p_\vartheta) = \frac{p_l^2}{2m}+ \frac{p_\vartheta^2}{2ml^2}+ \frac{1}{2}k \left( l - l_0 \right)^2 - m g l \cos \vartheta
\end{equation}
Making use of the same characteristic time and length scales, the following dimensionless quantities are defined:
\begin{equation}
	q_{\textnormal{p}1} = \frac{l}{l_0}, \quad 
	q_{\textnormal{p}2} = \vartheta ,\quad	
	p_{\textnormal{p}1} = \frac{p_l}{m l_0 \omega_S}, \quad 
	p_{\textnormal{p}2} = \frac{p_\vartheta}{m l_0^2 \omega_S}
\end{equation}
By applying this scaling to the Hamiltonian, we obtain the dimensionless form:
\begin{equation}
	\mathcal{H} = \frac{p_{\textnormal{p}1}^2}{2} + \frac{p_{\textnormal{p}2}^2}{2q_{\textnormal{p}1}^2} + \frac{1}{2}(q_{\textnormal{p}1} - 1)^2 - \omega^2 q_{\textnormal{p}1} \cos q_{\textnormal{p}2}
\end{equation}
The gravitational potential term $-\omega^2 q_{\textnormal{p}1} \cos q_{\textnormal{p}2}$ remains consistent with the Cartesian formulation, where the potential is minimized at the downward vertical ($q_{\textnormal{p}2} = 0$).

The corresponding dimensionless equations of motion in polar coordinates are derived as:
\begin{equation}
	\left\{
	\begin{aligned}
		\dot{q}_{\textnormal{p}1} &= p_{\textnormal{p}1} \\
		\dot{q}_{\textnormal{p}2} &= \frac{p_{\textnormal{p}2}}{q_{\textnormal{p}1}^2} \\
		\dot{p}_{\textnormal{p}1} &= \frac{p_{\textnormal{p}2}^2}{q_{\textnormal{p}1}^3} - (q_{\textnormal{p}1} - 1) + \omega^2 \cos q_{\textnormal{p}2} \\
		\dot{p}_{\textnormal{p}2} &= -\omega^2 q_{\textnormal{p}1} \sin q_{\textnormal{p}2}
	\end{aligned}
	\right.
	\label{Eq:motionPolar}
\end{equation}
Note that the two formulations (Eq.(\ref{Eq:motionPolar}) and Eq.(\ref{Eq:motionCart})) are related via the following point transformation in configuration space:
\begin{equation}
	\begin{aligned}
		q_{\textnormal{c}1} &= q_{\textnormal{p}1} \sin q_{\textnormal{p}2} \\
		q_{\textnormal{c}2} &= q_{\textnormal{p}1} \cos q_{\textnormal{p}2}
	\end{aligned}
	\label{Eq:TrafoPolarCard1}
\end{equation}
and the associated covariant transformation of the momenta
\begin{equation}
	\begin{aligned}
		p_{\textnormal{c}1} &= p_{\textnormal{p}1} \sin q_{\textnormal{p}2} + \frac{p_{\textnormal{p}2}}{q_{\textnormal{p}1}} \cos q_{\textnormal{p}2} \\
	p_{\textnormal{c}2} &= p_{\textnormal{p}1} \cos q_{\textnormal{p}2} - \frac{p_{\textnormal{p}2}}{q_{\textnormal{p}1}} \sin q_{\textnormal{p}2}
	\end{aligned}
	\label{Eq:TrafoPolarCard2}
\end{equation}
Conversely, the polar variables  can be expressed in terms of the Carthesian ones as:
\begin{equation}
	\begin{aligned}
		q_{\textnormal{p}1} &= \sqrt{q_{\textnormal{c}1}^2 + q_{\textnormal{c}2}^2} \\
		q_{\textnormal{p}2} &= \operatorname{atan}(q_{\textnormal{c}1}, q_{\textnormal{c}2})\\
		p_{\textnormal{p}1} &= p_{\textnormal{c}1} \sin q_{\textnormal{p}2} + p_{\textnormal{c}2} \cos q_{\textnormal{p}2} \\
		p_{\textnormal{p}2} &= q_{\textnormal{p}1} (p_{\textnormal{c}1} \cos q_{\textnormal{p}2} - p_{\textnormal{c}2} \sin q_{\textnormal{p}2})
		\label{Eq:TrafoPolarCard3}
	\end{aligned}
\end{equation}
This formal equivalence ensures that both formulations describe the same physical state in phase space. 
It must be noted however, that upon computational treatment numerical errors will refelct as different artifacts, eventually making one the superior choice of the other.
Moreover and more imporatant, as will be discussed in the following sections, the choice of the observation plane for Poincaré sections is often biased by the coordinate system used, where "straight" lines in one representation correspond to complex curves in the other. 
Recognizing this transformation is therefor key for developing a coordinate-independent methodology for phase-space analysis.

\subsection{Naive Poincaré Sections}
The phase space of a two-dimensional Hamiltonian system, such as the elastic pendulum, is four-dimensional. However, for an autonomous system, the total energy is a conserved quantity, restricting the motion to a three-dimensional isoenergetic hypersurface defined by $\mathcal{H}(\mathbf{q}, \mathbf{p}) = E$. On this manifold, one  variable (e.g., $p_2$) is implicitly determined by the other variables and the fixed energy value, i.e., $p_2 = p_2(q_1, q_2, p_1 | E)$. 
In this reduced space, a trajectory is represented by a continuous curve.

To visualize the global structure of such a 3D manifold, the method of Poincaré sections is commonly employed. A Poincaré section is a lower-dimensional subspace (a 2D surface) within the isoenergetic manifold. 
The dynamics are analyzed by observing the points where the trajectory pierces this surface in a specific direction (e.g., with a non-negative velocity component).
Formally, the choice of the piercing condition and the cross-section plane is arbitrary. 
However, the most frequent, and often naive, application involves setting one coordinate to zero, such as $q_i = 0$ with $p_i \geq 0$, and plotting the remaining momentum $p_j$ against its conjugate coordinate $q_j$. 
It is crucial to recognize the geometric implication of the choice $q_j = 0$ in configuration space. 
Different coordinate systems lead to different piercing conditions that correspond to distinct curves or lines in the physical space. 

\begin{figure}[bt]
	\centering
	\includegraphics[width=0.99\textwidth]{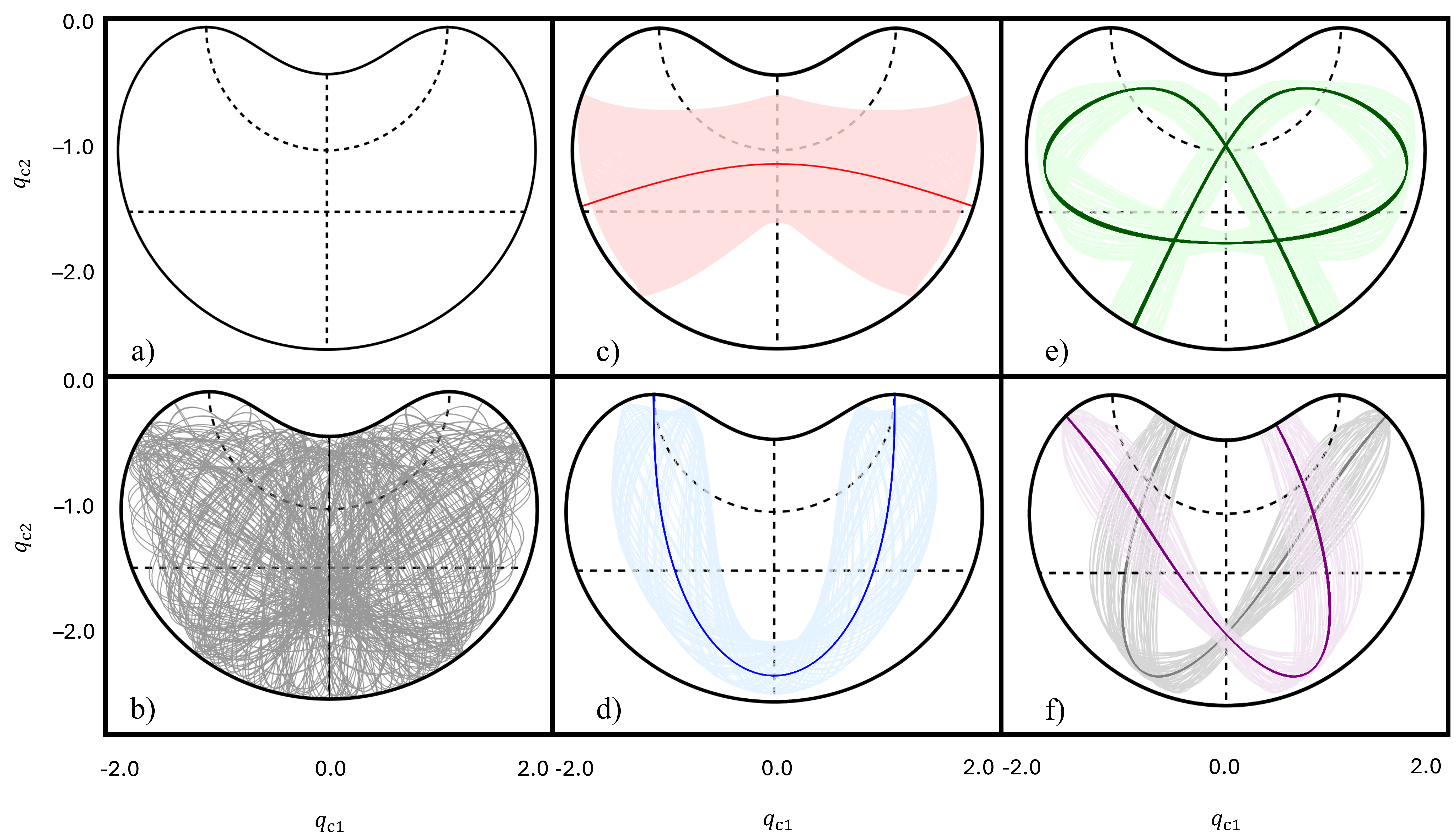}
	\caption{Overview of the fundamental periodic trajectories for $\omega^2=0.5$ and $E=0$: (b) unstable vertical oscillation, (c, d) stable symmetric cup and cap orbits, (e) asymmetric orbits, and (f) higher-order loop trajectory.}
	\label{Fig:naivSection_Trjectories}
\end{figure}

\Fref{Fig:naivSection_Trjectories}a displays the physical meaning of those naiv sections for the elastic pendulum.
In the Cartesian formulation (Eq. \eqref{eq:coord_cart}), the  choice  $q_{\textnormal{c}1} = 0$ corresponds to a vertical cut passing through the pivot point as well as the equilibrium. 
This section represents the symmetry line of the system.
On the other hand, the choice $q_{\textnormal{c}2} = 0$ refers to a horizontal section passing exactly through the pivot point. 
Since the equilibrium position of the mass lies at $q_{\textnormal{c}2} < 0$ and the associated equilibrium energy is $\mathcal H < 0$, this section is only accessible for high-energy cases where the mass can swing above the suspension point. 
Despite this limitation, such sections have been considered in studies focusing on high-energy bifurcations, such as the work by Anurag et al. \cite{anurag2022}. 
However, the definition of the vertical coordinate varies across the literature. 
While  Acosta-Zamora \textit{et al. \cite{AcostaZamora2024}} define the origin at the pivot point, other works such as H.~N. N{\'u}{\~n}ez-Y{\'e}pez and co-workers \cite{NUNEZYEPEZ1990101} effectively shift the origin or the reference length to the equilibrium point. 
For these works, a naive section defined by the "vertical" coordinate being zero actually resamples the horizontal line passing through the static equilibrium position, i.e., at a distance $q_{\textnormal{c}2} = y_{eq}/l_0 = -1 - \omega^2$.

In polar coordinates, the naive section defined by $q_{\textnormal{p}2} = 0$  also refers to the vertical symmetry line. 
In this specific case, the canaonical trasformation (Eq. \eqref{Eq:TrafoPolarCard3}) between the two formulation is actually a identical mapping with exception of the sign.
A significantly different section arises when considering the circular coordinate $q_{\textnormal{p}1}=0$. 
A section defined by $q_{\textnormal{p}1} = 0$ is physically meaningless as it represents the singular point at the pivot. 
However, when the coordinate is defined relative to the spring stretching---either as the natural length $l_0$  or relative to the equilibrium length---the naive section $q_{\textnormal{p}1} = \text{const.}$ reflects a circle centered at the pivot point. 
As such, the section condition $q_{\textnormal{p}1} = 1$ has been for example considered by Anurag {\it et al.} \cite{anurag2022}.

\section{Numerical simulation and Trajectory characterization}

\subsection{Computational Details}

Trajectories for  given initial conditions were obtained by numerically integrating the dimensionless Cartesian equations of motion (Eq.~\ref{Eq:motionCart}) utilizing the Radau~IIA9 implicit Runge-Kutta scheme with a constant integration step size of $10^{-3}$. \cite{hairer1999stiff,rackauckas2017differentialequations} 
To ensure numerical consistency, the system was additionally formulated in dimensionless polar coordinates (Eq.~\ref{Eq:motionPolar}), and the results obtained from both formulations were found to be in excellent agreement within numerical accuracy upon applying the canonical transformations defined in Eqs. (\ref{Eq:TrafoPolarCard1}---\ref{Eq:TrafoPolarCard3}). 
Poincaré mappings were generated by evaluating the specific sectioning condition in Cartesian coordinates. 
To locate the precise piercing points, the trajectory data points surrounding a crossing were interpolated via higher-order Lagrange polynomials, and a bisection root-finding algorithm was subsequently employed to resolve the intersection point up to a target tolerance of $10^{-12}$.
 For customized section criteria that were not available analytically but defined numerically (see Sections~\ref{SubSec:ForceSections} and \ref{SubSec:TrajSections}), the geometry of the sectioning surface was reconstructed via linear interpolation between pre-computed neighboring pivot points.

For the practical implementation of the force-line and isopotential coordinate systems (see Section~\ref{SubSec:ForceSections}), a continuous numerical scheme was deployed. 
Initially, the arc length along the specific isopotential boundary line ($V(x,y) = V_0=\mathcal H$) was evaluated via numerical quadrature using Simpson's rule, integrating from the lower intersection of the potential curve with the $y$-axis up to the desired spatial coordinate. 
Taking this coordinate as the initial condition, the corresponding force field line was subsequently generated by numerically integrating the normalized negative potential gradient in Cartesian coordinates using the adaptive 9th-order Verner Runge-Kutta algorithm (\texttt{Vern9}). \cite{verner2010numerically,rackauckas2017differentialequations}
Finally, the components of the Jacobian matrix at any given pivot point along the generated force field line was determined via finite differences, considering neighboring force lines separated by a circumferential distance of $10^{-4}$, whereby the fields of neighboring force lines were linearly interpolated for spatial locations falling between explicit pivot points.

To construct the Frenet-Serret frame along the defined stable reference trajectories (see section \ref{SubSec:TrajSections}), a single closed periodic orbit in the Cartesian configuration space was extracted from the numerical solution of the equations of motion. 
These discrete trajectory points were then fitted using third-order cubic splines to ensure differentiable geometric continuity. 
Upon detecting the intersection of an arbitrary trajectory with this section, the Jacobian matrix required to canonically transform the conjugated momenta into the local Frenet-Serret frame was directly computed from the analytical derivatives of the fitted cubic splines at the exact point of intersection.

\subsection{Categorization of Trajectories}

To provide a systematic evaluation of the mappings obtained through the different sections discussed above, it is first necessary to characterize the fundamental types of trajectories that occur in the elastic pendulum. 
This identification and classification of these orbits are crucial, as they provide the reference points needed to judge whether a given Poincaré section accurately captures the global topology or introduces misleading geometric artifacts.

In this study, we focus on the specific case where $\omega^2=0.5$ and the dimensionless energy is fixed at $\mathcal H=0$. 
At this energy level, the phase space exhibits a well-balanced coexistence of chaotic and regular regions, making it an ideal candidate for comparing the topological clarity of various Poincaré sections. While we have performed extensive studies on how the dynamics depend on these parameters---specifically regarding the transitions and bifurcations that manifest in individual sections---such a parametric analysis lies beyond the scope of the present work and will be reported elsewhere.

Within this specific parameter regime, the phase space "skeleton" is defined by five fundamental periodic orbits, which appear as fixed points in a suitably chosen Poincaré map. 
The first of these is the pure vertical oscillation (\fref{Fig:naivSection_Trjectories}b), representing the motion where the system remains entirely on the symmetry axis. 
In this regime, the vertical oscillation acts as an unstable fixed point.
This instability is a consequence of the concave curvature of the potential energy surface at the upper reaches of the motion; any infinitesimal lateral perturbation is amplified as the mass "falls off" the potential ridge.
 This mechanism is the physical driver behind the autoparametric resonance frequently cited in literature, where energy is transferred from the radial spring mode to the angular pendulum mode, typically resulting in a wide chaotic sea surrounding the vertical solution.

In contrast to the unstable vertical motion, the system possesses stable symmetric fixed-point-one trajectories.
 These orbits maintain the reflection symmetry of the Hamiltonian and are characterized by their distinct geometries in configuration space: the upward-opening "cup" ($\cup$) and the downward-opening "cap" ($\cap$) (\fref{Fig:naivSection_Trjectories}c, d).
 As stable fixed points, they serve as the centers of invariant tori, meaning that nearby trajectories oscillate quasi-periodically around them. 
 It is important to note that these symmetric orbits cross the vertical symmetry line perpendicularly, ensuring that the path on the right-hand side is a precise mirror image of the path on the left. 
Furthermore, the system exhibits asymmetric fixed-point-one trajectories (\fref{Fig:naivSection_Trjectories}e). Unlike the cup and cap orbits, these paths cross the symmetry line diagonally, resulting in a trajectory that is not self-mirroring.
 Due to the underlying symmetry of the problem, these asymmetric orbits always exist as degenerate pairs—mirror images of one another that share the same energy and stability characteristics. 
Finally, the structure is completed by  higher-order fixed-point trajectories, such as the "loop" orbit often depicted in \fref{Fig:naivSection_Trjectories}f).

\section{Naiv Poincaré sections}

A consideration of several trajectories obtained from random inisilaisation in the isoenergetic subspace of the phacespace is now considered for the naiv poincaré section presented above. 
The obtained mappings are presented in \fref{Fig:naivPoincareMaps}, where the coloring of individual trajectories are acording to the types of trajectories: cap islands red, cup islands blue, loop-island green, chaotic gray, asymetric purple.

\subsection{vertical section}

\begin{figure}[bt]
	\centering
	\includegraphics[width=0.99\textwidth]{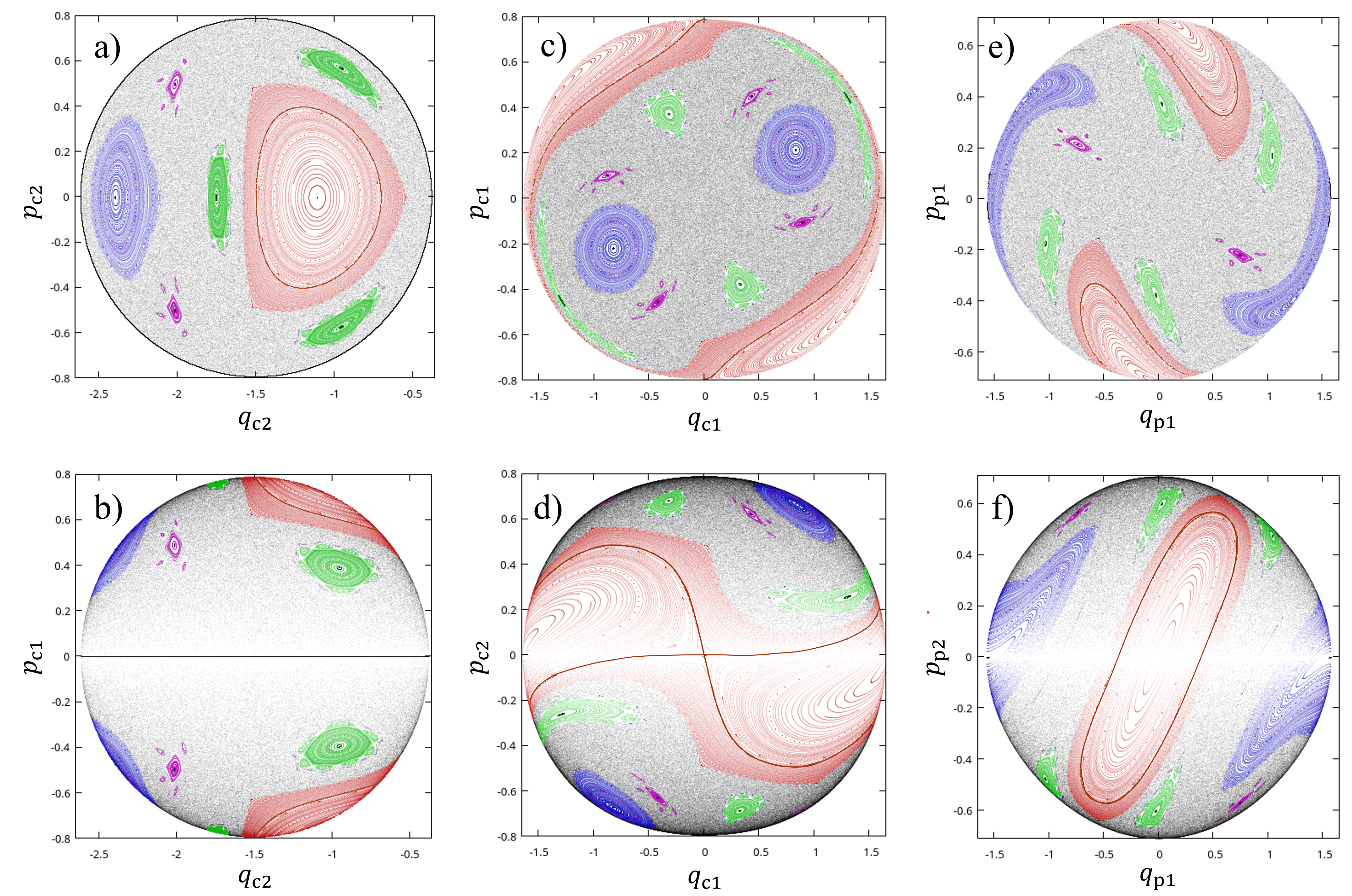}
	\caption{Traditional and inverted naive Poincaré sections illustrating the global phase-space organization. Trajectories are color-coded as follows: cap islands (red), cup islands (blue), loop islands (green), chaotic sea (gray), and asymmetric islands (purple). Panels display: (a) the traditional vertical section at $q_{c1}=0$, (c) the traditional horizontal section passing through the static equilibrium at $q_{c2}=-1-\omega^2$, and (d) the traditional radial section at the unstretched natural length $q_{p1}=1$. Panels (b), (d), and (f) show the corresponding inverted mappings for the vertical, horizontal, and radial sections, respectively, plotting the orthogonal momenta to reveal hidden topological transformations such as apparent separatrices and centered quasi-periodic stability islands.}
	\label{Fig:naivPoincareMaps}
\end{figure}

We begin our comparative analysis by examining the results obtained from the vertical section defined by $q_{\textnormal{c}1}=0$ (or equivalently $q_{\textnormal{p}2}=0$), as shown in \fref{Fig:naivPoincareMaps}a. 
A defining feature of this specific section is its mirror symetry with respect to the horicontal axis.
Another propertiy is that it contains an entire trajectory within its own plane: the vertical harmonic oscillation.
For this unique solution, the momentum component perpendicular to the section ($p_{\textnormal{c}1}$) remains zero at all times. 
Consequently, the vertical oscillation serves as the absolute boundary of the accessible phase space in this representation, manifesting as an elliptical  curve that encapsulates the map.
The interior of the map is dominated by an extensive chaotic sea indicated in gray. 
Notably, this chaotic region directly approaches the elliptical boundary across its entire perimeter.
This visual proximity confirms the previously mentioned association of caotic trajectories as pertubation of the vertical harmonic motion. 
Embedded within this chaotic sea are the stability islands of the regular orbits.
The prominent blue and red structures represent the symmetric "cup" and "cap" orbits, respectively. 
Surrounding these fixed points are the invariant tori of quasi-periodic motion. 
Additionally, the higher-order fixed points, such as the fixed-point-three "loop" trajectory (green) and the asymmetric fixed-point-one pairs (purple), are visible as distinct islands.

\subsection{horizontal section}

A markedly different perspective of the phase-space topology is obtained by considering the horizontal section passing through the static equilibrium position, defined by the condition $q_{\textnormal{c}2} = -1-\omega^2$. As shown in \fref{Fig:naivSection_Trjectories}c), the visual arrangement of the stability islands and the chaotic sea undergoes a profound transformation.
Instead of a mirrow symetry with respect to the horizontal axis, the mapping has now a inflection symetry with respect to its  center.
This is a consequnce of the fact, that the section is perpendicular to the axis of the system.
Moreover, the red and blue  islands corresponding to the "cap" and "cup" orbits no longer appear ones as centered regions but  twice.
This is due to the fact that the orbits pass the section on the left and the right equally. 
The quasi-periodic orbits surrounding the cap orbit (red) are shifted to the periphery of the  mapping, appearing as elongated, crescent-shaped structures reaching to the  boundaries of the map.
The fix-point of the cap trajectory itself, however, is not present in the section since the cap orbit remains entirely above the section, see  \fref{Fig:naivSection_Trjectories}c).
Trajectories adjacent to the cap orbit, on the other hand, are pierced by the horizontal section at grazing angles near their maximum lateral extension where the moment is largely parallel to the section and therefor located near the boundary of the mapping. 
This causes the strong deformation of these specific tori.  
In contrast, the cup trajectory reaches its maximum downward extension well below the equilibrium line, see Figure. 
Consequently, each full cycle of the orbit intersects the horizontal line $q_2=-1- \omega^2$ in  resonable steep angles, maintaining the associated region in the mapping apart from the boundary. 

The regions that originate from the assymetric fix-point-1 orbit also doubles and appear again as satelite of the cup trajectories.
The region associated with the loop trajectory on the other hand appears now four times, rather than three as observed in the vertical section map. 
Whereas these features previously appeared as satellite structures surrounding the cap region, they now manifest as independent structures.
Two of the four islands are located in a highly stretched region near the map's boundary, appearing as thin, elongated structures. 
This extreme deformation reflects the fact that the horizontal section pierces the associated tori at a near-tangential angle at their lateral extrema. 
In contrast, the remaining two islands are situated in the interior of the map and retain a more regular, compact shape. 
This discrepancy in appearance is again due to the fact, that the piecring appears under a agreacing angle for one part of the orbit and more steep for another, compare \fref{Fig:naivSection_Trjectories}.

The chaotic sea (gray) remains a dominant feature, but its boundary is now confioned by the quasi-periodi orbita associated to the cap region.
Concluding the presentation of the naiv horizontal section, we comment that although presenting an overall eliptical shape, the envelope of the map does not resample a special shape, and in particular no pure elipse.

\subsection{Radial Section}

The third naive section arrises from the use of the polar formulation and by defining a radial section at the spring's unstretched natural length, $q_{\textnormal{p}1} = 1$ (with $p_{\textnormal{p}1} \geq 0$). 
The resulting Poincaré map, spanned by the angular coordinate $q_{\textnormal{p}2}$ and its conjugate angular momentum $p_{\textnormal{p}2}$, is presented in \fref{Fig:naivPoincareMaps}c.
A coresponding mapping for a section considering the equilibrium length $q_{\textnormal{p}1} = 1+\omega^2$ is presented and discussed in detail in teh Supplemantary section X.

Similar to the horizontal cut, this map exhibits an inversion symmetry with respect to the center rather than a mirror symmetry which again  reflects the symetry of  the system.
The geometric projection of this radial section again drastically alters the representation of the primary regular orbits. 
The cap orbit (red),  is now split into two distinct half-islands located at the top and bottom  of the map.
 Because the cap trajectory largely oscillates around a stretched configuration, it pierces the $l=l_0$ plane at high radial velocities, resulting in a diagonal stretching in the angular phase space. 
As before in the horizontal section, the cap trajectory itself is not pierced by the section such that only its quasi-periodic stability island are part of the mapping.

Even more extreme is the deformation of the cup orbit (blue). 
In this radial representation, the cup islands are pushed to the absolute periphery of the accessible phase space, manifesting as thin, severely distorted slivers along the map's boundary. 
This peripheral localization occurs because the cup trajectory's radial extension takes place predominantly at larger lengths.
As noticible in \fref{Fig:naivSection_Trjectories}d) the cup trajectory does barely grazes the section $l=l_0$ radius  meaning it is also not part of the map

The higher-order resonances also adapt to this circular geometry. 
The region associated with the loop trajectory (green) again appears as four independent islands, but they are now stretched diagonally, aligning with the "flow" of the adjacent cap regions, making them appear as satailes.
The asymmetric fixed-point-one pairs (purple) remain situated in the interior but appear entirely only twice instead of four times as in the horizontal section. 
Only a small part of the outmost stability island is present near the boundary of the section.

Finally, the bounding envelope of the accessible phase space in this section has again no specific shape. 
It is defined by the energy limit where the radial kinetic energy is zero ($p_{\textnormal{p}1} = 0$). Evaluating the Hamiltonian under this condition yields the restriction $p_{\textnormal{p}2}^2 = 2(E + \omega^2 \cos q_{\textnormal{p}2})$. This purely angular energy bound dictates the eye-shaped boundary visible in the figure. 

\subsection{Inverted Mappings}

The traditional mapping which plots the momentum $p_j$ against its conjugate coordinate $q_j$ (subject to the piercing condition $p_i \ge 0$), tents to present the phase space distorted as trajectories at the boundary of the map are strongly squeezed in comparison to regions in the center of the plot. 
This may suppress the visibility of some features of the mappings, particularly those that are consequence of the grazing angle of the section.
To better reveal those properties, it is meaningful to also consider the inverted mappings where the exact same geometric cut but plots the orthogonal momentum $p_i$ against $q_j$ under the complementary condition $p_j \ge 0$ is mapped.
These corresponding maps for the vertical, horizontal end radial sections are presented in \fref{Fig:naivPoincareMaps} d,e and f, respectivally.

To understand the topological relationship between the traditional and inverted mappings, one must consider the constraints imposed by energy conservation. 
For a conservative system with a fixed total energy $E$, the state of the system on the sectioning plane ($q_i=0$) dictates that the potential energy $V(q_i=0, q_j)$ depends exclusively on the coordinate $q_j$. Consequently, for any specific point $q_j$ on the horizontal axis of the map, there is a strictly defined amount of kinetic energy available, which must be shared between the relevant momenta. 
Because the kinetic energy scales with the squares of the momenta, there is a direct trade-off: an increase in $p_j^2$ strictly requires a proportional decrease in $p_i^2$, and vice versa.
This energetic balance leads to a profound topological transformation where the inverted map effectively cuts the traditional mapping in half and turns it "inside out." 
Specifically, the outer boundary of a traditional $p_j$-versus-$q_j$ plot represents the energetic limit where the plotted momentum $p_j$ is maximized, meaning the hidden momentum component $p_i$ approaches zero. 
In the inverted plot ($p_i$ versus $q_j$), this exact boundary state maps directly to the horizontal axis where $p_i=0$. 
Conversely, the horizontal axis of the traditional map ($p_j=0$) corresponds to the states where the system possesses maximum $p_i$, which consequently defines the outer envelope of the inverted map.

Looking at the vertical sections, for instance, the prominent symmetric "cap" orbit (red) that occupies the central region of the traditional map (\fref{Fig:naivPoincareMaps}a) is pushed to the extreme upper and lower boundaries in the inverted map (\fref{Fig:naivPoincareMaps}d). Similarly, the absolute boundary of the traditional vertical map---which represents the pure vertical harmonic oscillation where $p_{\textnormal{c}1}=0$---collapses exactly onto the horizontal axis in the inverted representation.

This pattern of inversion is consistently observed across the different section choices, yet it reveals unique topological features that are otherwise obscured. 
In the inverted horizontal section (\fref{Fig:naivPoincareMaps}e), the "cap" trajectories (red) exhibit a striking geometric departure from the nested island structures seen in all other mappings.
 While the traditional maps typically show these islands as concentric, encapsulated regions, the inverted horizontal cut reveals a shape resembling an avoided crossing in the center of the map. 
The regions appear to be partitioned by a distinct trajectory (marked dark red) that acts as an apparent local separatrix.
Interestingly, a comparison with the other mappings shows that this specific trajectory behaves similarly to the other  quasi-periodic orbits of the "cap" family; its unique appearance here is not an intrinsic dynamical change, but rather a projection artifact where the sectioning plane pierces the torus at a singular angle where the kinetic energy distribution between $p_{\textnormal{c}1}$ and $p_{\textnormal{c}2}$ undergoes a reversal.

The inverted radial section (\fref{Fig:naivPoincareMaps}f) provides a similarly transformative view. In the traditional radial map (\fref{Fig:naivPoincareMaps}c), the "cap" islands were split and pushed toward the boundaries. 
In the inverted representation, however, the "cap" region appears centered as a single, coherent region, providing a much clearer view of its internal stability. 
Simultaneously, the "cup" region (blue), which was reduced to nearly invisible slivers in the traditional radial cut, is strongly enlarged in the inverted plot. 
This expansion effectively "zooms in" on the cup's stability tori, allowing for a detailed visualization of its features that were previously hidden due to the tangential nature of the traditional radial crossing. 

Ultimately, these comparisons visually confirm that the inverted sections capture the exact same dynamical structures but project them through a reversed kinetic lens, occasionally resolving features that their traditional counterparts fail to display clearly.

\section{Construction of New Sections and associated coordinates}

Among the traditional sections discussed so far, the vertical section stands out due to two inherent properties. First, at each point along this section, the net force is parallel to the section plane, meaning the section acts as a force line. Second, as mentioned previously, there is a specific trajectory that takes place entirely within this section: the pure vertical harmonic oscillation. 
We now use these two physical properties individually to define new sections. Our goal is to analyze the behavior of Poincaré mappings when they are tied to these specific properties of subspaces of the phase space. 

Simply implementing these new sectioning conditions using cartesian or polar coordinates is highly inconvenient. 
Defining a curved section as $S(q_1, q_2, p_1, p_2) = 0$ means that none of the  variables $(q_1, q_2, p_1, p_2)$ remain constant along the cut. As a result, generating a complete visualization or the poincaré maps by plotting $p_j$ over $q_i$ would require four individual plots. 
Moreover, the section condition is only a necessary condition to obtain the mapping. 
For a unique representation, we also need a sufficient criterion that guarantees a constant piercing direction through the section. Simply choosing $p_i > 0$ does not provide this guarantee when the section itself is curved. 
It is therefore meaningful to choose tailored coordinates, such that these desired new sections are again represented as naive sections, i.e., by choosing $q_i = \text{constant}$. 
By defining the proper conjugated momenta for these new coordinates, the sufficient condition for visualizing the Poincaré maps is then simply given by $p_i > 0$ again.

\subsection{Forceline section}
\label{SubSec:ForceSections}

We start by constructing a orthogonal set of coordintes $(q_{\text{f}1},q_{\text{f}2})$ such that each force line is specified as $q_{\text{f}1}= \text{constant}$. 
Due to the inherent propery of force lines and isopotential lines to be othogonal, the second coordinate $q_{\text{f}2}$ must be a function of the potential $V$, which provides a direct transformation equation from the traditional coordinates $(q_{\text{c}1},q_{\text{c}2})$ or $(q_{\text{p}1},q_{\text{p}2})$, respectivally.
To uniquely determine the gauge of $q_{\text{f}2}$,  the transformation is requzired to be regular and free of coordinate-induced singularities at the elliptic equilibrium.
For an arbitrary parameterization $q_{\text{f}2} = f(V - V_{\text{min}})$, the conjugate momentum $p_{\text{f}2}$ scales asymptotically near the minimum as
\begin{equation}
	p_{\text{f}2} \propto \frac{1}{\sqrt{V - V_{\text{min}}} \cdot f'(V - V_{\text{min}})}\,.
\end{equation}
This universal scaling is a direct consequence of the invariance of the symplectic form. Since the physical configuration space area enclosed by a near-equilibrium quadratic energy contour scales as $\text{Area} \propto \sqrt{V - V_{\text{min}}}$, any canonical transformation requires the conjugate momentum space to scale inversely to the spatial deformation to preserve a regular, non-vanishing volume of the underlying invariant tori \cite{arnold1989mathematical}. Consequently, the choice 
$$q_{\text{f}2} = \sqrt{V - V_{\text{min}}}$$
represents the unique gauge that ensures a regular and bounded momentum boundary $p_{\text{f}2} \to p_0 > 0$ as $q_{\text{f}2} \to 0$.


\begin{figure}[!bt]
	\centering
	\includegraphics[width=0.99\textwidth]{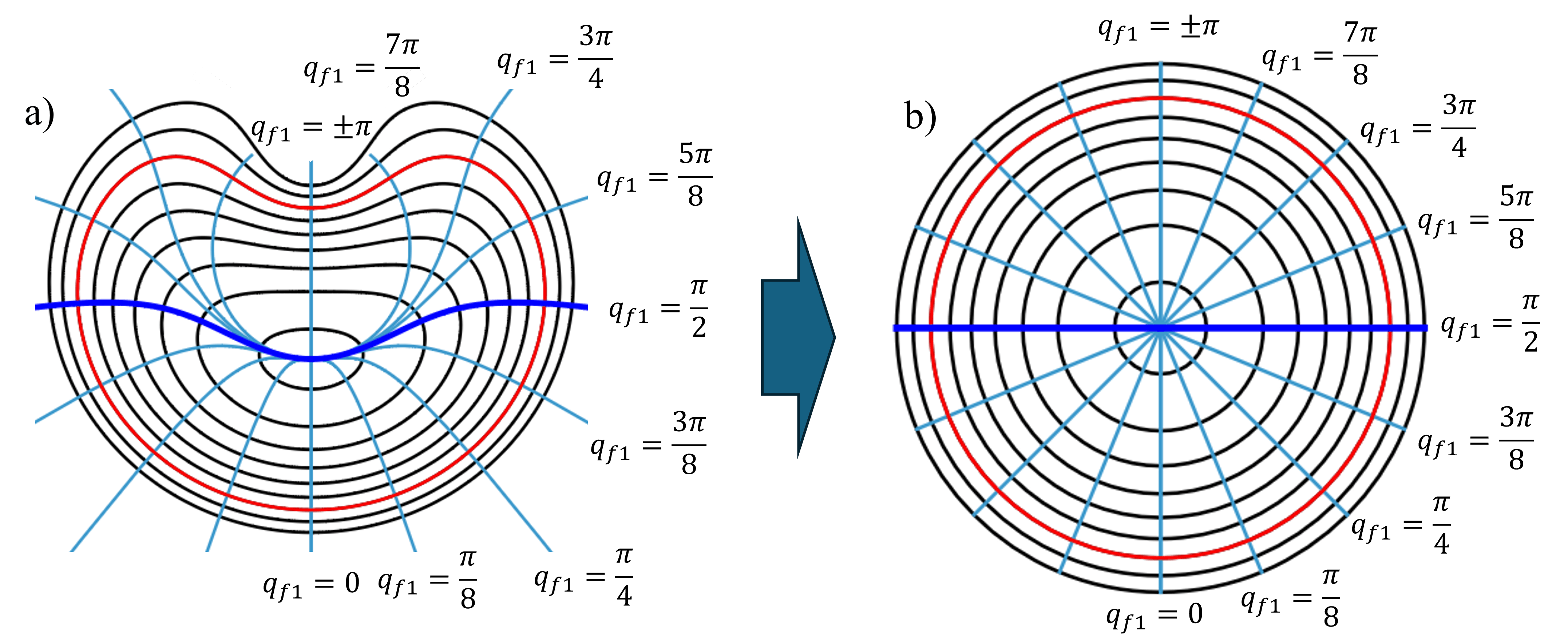}
	\caption{Construction of the force-line and isopotential coordinate system $(q_{f1}, q_{f2})$. Panel (a) shows the definition of the force-line parameter $q_{f1}$ via the normalized arc length along the isoenergetic isopotential boundary $\mathcal{H}=0$ in Cartesian configuration space. Panel (b) illustrates the resulting deformation of the configuration space where the new coordinates act as polar configurations, with force lines becoming infinitely densely packed as they approach the equilibrium.}
	\label{Fig:TrafoForceCoords}
\end{figure}

While the definition of $q_{\textnormal{f}2}$ is analytically accessible through the potential energy, the definition of the force-line parameter $q_{\textnormal{f}1}$ is less intuitive. 
In fact, an explicit analytical transformation that uniquely indexes each force line is not directly available. This is because indexing the family of gradient curves analytically requires finding a closed-form first integral of the non-linear gradient flow equations $\dot{\mathbf{r}} = -\nabla V(x,y)$. 
Due to the non-linear coupling of degrees of freedom via the spring-stretching term $\sqrt{x^2+y^2}$ in the pendulum's potential, this gradient system is non-separable and does not possess a known algebraic invariant. 

While the definition of $q_{\textnormal{f}2}$ is analytically accessible through the potential energy, the definition of the force-line parameter $q_{\textnormal{f}1}$ is less intuitive. In fact, an explicit analytical transformation that uniquely indexes each force line is not directly available. This is because indexing the family of gradient curves analytically requires finding a closed-form first integral of the non-linear gradient flow equations $\dot{\mathbf{r}} = -\nabla V(x,y)$. Due to the non-linear coupling of degrees of freedom via the spring-stretching term $\sqrt{x^2+y^2}$ in the pendulum's potential, this gradient system is non-separable and does not possess a known algebraic invariant.

To overcome this fundamental limitation, this study relies on a numerical assessment utilizing the following approach. 
Starting from any arbitrary point, the unique line of force passing through it can be determined by tracking the gradient of the potential via numerical integration.
To map these lines onto a well-defined coordinate $q_{\textnormal{f}1}$, we consider the arc length along the isoenergetic isopotential line in the cartesian configuration space with energy equivalent to the considered value of the Hamilton function ($\mathcal{H}=0$ in our case). 
To ensure a standardized and convenient representation, we normalize this total arc length to the interval $[-\pi, \pi]$, where $q_{\textnormal{f}1}=0$ refers to the lower intersection of the isopotential line with the $y$-axis, see \fref{Fig:TrafoForceCoords}a.
Since each force line approaches the equilibrium point while the isopotential lines maintain concentric around it, the thus obtained transformation can be understood as a deformation of the configuration space where the new coordinates act as polar configurations, see \fref{Fig:TrafoForceCoords}b.
Note, that as the force lines approach the equilibrium, they get infinitely densely packed. This means that at the equilibrium $q_{\textnormal{f}1}$ is not uniquely defined, reflecting a singularity in this set of coordinates.
In order to complete the canonical transformation to our new set of variables, the transformations for the conjugated momentum $p_{\textnormal{f}1}$ is required. Since in hamiltonian systems the momenta transform contravariantly, they can be obtained through the inverse of the Jacobian matrix. Because the coordinate $q_{\textnormal{f}1}$ is only available numerically, the partial derivatives composing the Jacobian are also assessed numerically through a finite difference approximation. 

Remarkably, despite the numerical nature of this grid, the asymptotic behavior of the tracking momentum can be predicted from the local anisotropy of the potential minimum. Near the elliptic fixed point, the gradient flow $\dot{\mathbf{r}} = -\nabla V$ becomes asymptotically homogeneous, forcing the geometric pathways of generic force lines to scale as $\tilde{y} \propto x^\lambda$, where $\lambda = K_y/K_x = 1 + \frac{1}{\omega^2} > 1$ represents the ratio of the local stiffnesses \cite{guckenheimer2013nonlinear}. 
Because the covariant transformation inherits this anisotropic compression, the maximum available conjugate momentum along any generic line of force obeys the power-law
\begin{equation}
	p_{\textnormal{f}1,\text{max}} \propto q_{\textnormal{f}2}^\lambda \quad (\text{as } q_{\textnormal{f}2} \to 0).
\end{equation}
Due to the discrete symmetry of the system, this behavior exhibits a sharp discontinuity exactly on the vertical axis ($x=0$, corresponding to $q_{\textnormal{f}1} = 0, \pm\pi$). 
On this singular line, the spatial roles of the normal modes invert, causing the scaling to self-invert to $q_{\textnormal{f}2}^{1/c}$.

As indicated in \fref{Fig:TrafoForceCoords}, considering a Poincaré section through $q_{\textnormal{f}1}=\text{const.}$ would limit the considered instances to only one side of the potential. While due to symmetry this does not result in a lack of information, it is nevertheless helpful to consider both sides of the potential. This can be obtained by combining the instances $q_{\textnormal{f}1}=c$ and $q_{\textnormal{f}1}=c-\pi$. Note that in this way, the vertical section is restored by combining $q_{\textnormal{f}1}=0$ and $q_{\textnormal{f}1}=\pm\pi$.

After careful definition of a coordinate system in which individual force lines are presented as $q_{\text{f}1} = \text{constant}$, we now proceed to present and discuss the associated Poincaré maps. For this analysis, we focus on the section $q_{\text{f}1} = \pm\pi/2$, while maps for further section choices can be found in the Appendix. 
\fref{Fig:PoincareForce} presents the Poincaré mapping obtained from this force-line section. Globally, the representation shares similar properties with the traditional horizontal section (\fref{Fig:naivPoincareMaps}b+e), which is a direct consequence of the fact that the two sectioning surfaces remain geometrically close to each other within the configuration space. 
In particular, the number of islands for the individual types of trajectories remains identical. 
Key differences, however, emerge when examining the boundaries and the fine structure of the stable islands. Due to the curved nature of the section in the Cartesian configuration space, the enveloping shape of the map clearly deviates from an elliptical shape, taking on a distinct, hexagon-like appearance. This geometric deformation is rooted in the intrinsic properties of the force lines, which undergo a transition from convex to concave curvature. The presence of these inflection points along the sectioning curve alters the projection of the outer trajectories. Consequently, this causes the continuous regular frame at the periphery to break up, allowing the chaotic sea to extend further outward or interlock with fragmented stability islands in a hybrid boundary structure. 
Furthermore, the geometric stretching inherent to the force-line coordinates alters the apparent size and distribution of the primary resonance islands in the central region of the map, providing a more balanced view of the phase-space density near the equilibrium. In particular, the section now incorporates the stable period-one "cap" fixed point, which was not part of the horizontal map.

\begin{figure}[!bt]
	\centering
	\includegraphics[width=0.9\textwidth]{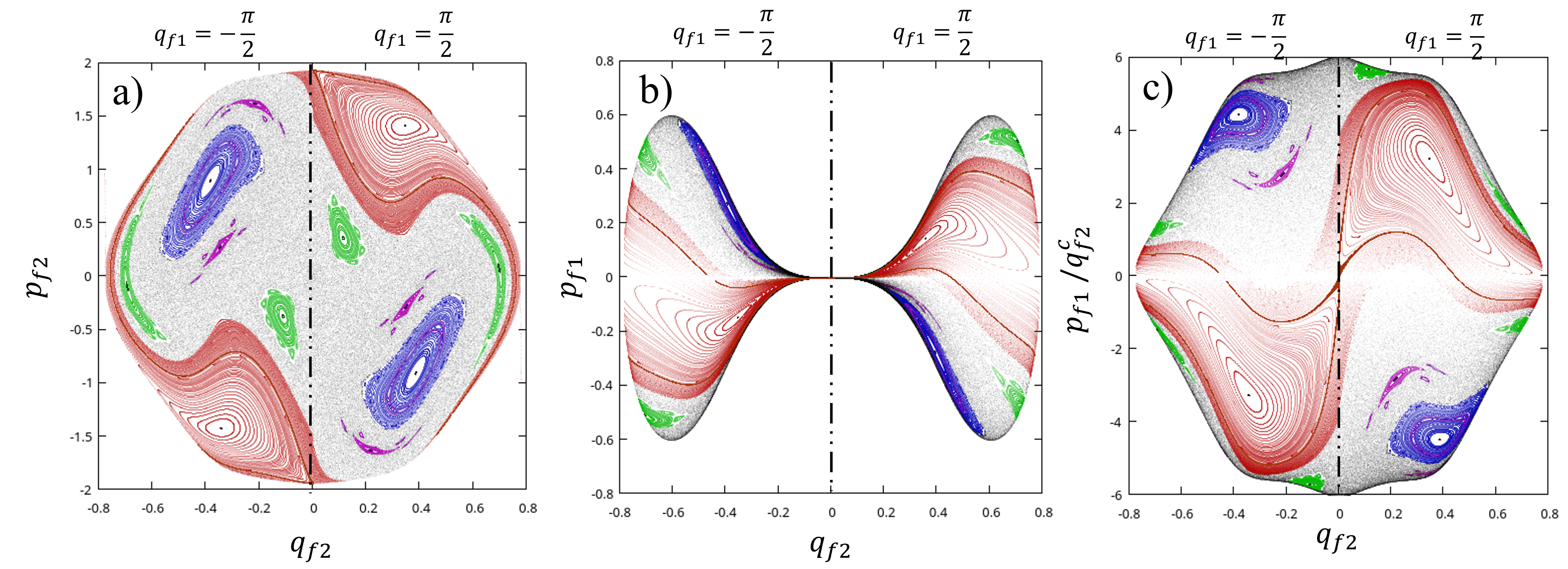}
	\caption{Poincaré mappings derived from the customized force-line section at $q_{f1}=\pm\pi/2$. Panel (a) shows the direct mapping, which exhibits a hexagon-like boundary deformation and successfully incorporates the stable period-one "cap" fixed point. Panel (b) displays the unscaled inverted mapping, highlighting the severe geometric confinement and pinching as the system approaches the elliptic equilibrium. Panel (c) presents the scaled inverted mapping where the renormalized momentum unfolds the central pinch. }
	\label{Fig:PoincareForce}
\end{figure}

Next, the inverted mapping of the force-line section is analyzed in detail, where the tracking momentum $p_{\textnormal{f}1}$ is plotted against the potential coordinate $q_{\textnormal{f}2}$ under the condition $p_{\textnormal{f}2} > 0$ (see \fref{Fig:PoincareForce}b). The most prominent features within this representation are the severe geometric confinement and the resulting compression near the center of the map. 
As the system approaches the elliptic equilibrium ($q_{\textnormal{f}2} \to 0$), the entire phase-space structure is pinched along the vertical axis, forcing the tracking momentum to vanish smoothly as $p_{\textnormal{f}1} \to 0$. 
This visual pinching provides direct numerical evidence for the anisotropic scaling law derived from the local gradient flow geometry above.
 While this severe compression masks the fine structure of the near-equilibrium invariant tori to the naked eye, it highlights the structural boundary constraints imposed by the coordinate transformation.
Beyond this spatial restriction, the inverted projection induces a significant topological translocation of the stability islands compared to both the traditional horizontal map and the direct map discussed above. Remarkably, the stable period-one "cap" island chain (shown in red), which typically occupies the peripheral or exterior regions of traditional representations, is now shifted deep into the interior of the lateral "wings". Conversely, the "cup" resonance structures are dislocated toward the outer boundaries of the mapping. Furthermore, the geometric distortion across the section sectors causes one of the major secondary loop island chains to split, rearranging the chaotic sea into a highly asymmetric distribution across the wings.

Leveraging the asymptotic behavior of $p_{\textnormal{f}1}$ near the equilibrium, a scaled inverted mapping is presented in \fref{Fig:PoincareForce}c, where the renormalized momentum $p_{\textnormal{f}1}/q_{\textnormal{f}2}^c$ (with $c=3$ for the parameters considered here) is plotted against $q_{\textnormal{f}2}$. 
The renormalization successfully "unfolds" the central pinch observed in the unscaled representation. 
For all generic lines of force, the invariant curves approach well-defined, finite non-zero limits as $q_{\textnormal{f}2} \to 0$, which provides a rigorous numerical verification of the power-law prediction $p_{\textnormal{f}1,\text{max}} \propto q_{\textnormal{f}2}^c$. 
This unfolding dramatically enhances the visibility of the interior phase space, uncovering fine chaotic layers and small-scale secondary island chains near the equilibrium that are completely masked in any unscaled projection. 
Furthermore, it allows for a direct, high-resolution comparison with the structural features of the traditional horizontal section.
In particular, the apparent separatrix is formed again, which is in line with the fact that near the equilibrium, the force-line section asymptotically approaches the horizontal section. 
For larger values of $q_{\textnormal{f}2}$, the mapping deviates more significantly due to the geometric deformation of the grid, yet remains topologically very similar; specifically, the "cap" and the asymmetric stability islands appear merely displaced and distorted. 
Key advantages, however, emerge regarding the accessibility of specific orbits: the force-line section now fully includes the stable period-one trajectory of the "cup" island. Unfolding this region allows one to directly visualize and evaluate subtle dynamical properties such as the winding number profile or the fractality of the surrounding chaotic sea. On the other hand, the prominent loop islands are pushed toward the outer boundaries and are divided between the upper and lower sectors of the map, illustrating how the adaptive coordinate system redistributes phase-space features to optimize the resolution of near-equilibrium dynamics.

The fundamental concept of constructing a canonical coordinate system based on the lines of force and isopotential lines elaborated here is by no means restricted to the system of the elastic pendulum.
Instead, this geometric approach can be generalized to any smooth potential, particularly in the vicinity of local minima. Such an intrinsic formulation is highly advantageous for complex systems that lack inherent spatial symmetries, where traditional rigid Cartesian or polar slices often inadvertently slice across invariant structures and distort or obscure crucial phase-space properties. By adopting the natural metric of the force field, the resulting maps inherently conform to the flow, and the underlying mathematical framework strongly suggests that if one sectioning surface is evaluated, its orthogonal counterpart should be considered as well to gain a complete understanding of the system's canonical structure.
It must be emphasized, however, that the procedural definition of the force-line indexing parameter is not uniquely fixed. While measuring the normalized arc length along the outermost isopotential boundary of the energetically confined domain serves as a robust and intuitive guideline for localized wells, alternative choices may prove advantageous for more complex landscapes.
 This is particularly relevant for multi-well potentials or fields featuring several stationary points and saddles. In such configurations, individual lines of force are no longer topologically bound to reach the outer energy boundary, but may instead terminate at saddles or diverge along asymptotic channels. For these advanced applications, adapting the normalization anchor to internal separatrices or using local field-line invariants can optimize the regularity of the transformation, paving the way for a highly flexible diagnostic tool in the global analysis of multi-degree-of-freedom Hamiltonian systems.

\subsection{Fix-point-one trajectory section}
\label{SubSec:TrajSections}

The second inherent feature of the traditional vertical section is that it aligns with a continuous fixed-point trajectory.
We now consider how phase-space properties are presented in a Poincaré mapping when this specific feature is used to define new sectioning surfaces.
A coordinate transformation for which these sections are given by setting one coordinate constant can be obtained by parameterizing the path in the configuration space described by the corresponding fixed-point trajectory. 

\begin{figure}[!bt]
	\centering
	\includegraphics[width=0.9\textwidth]{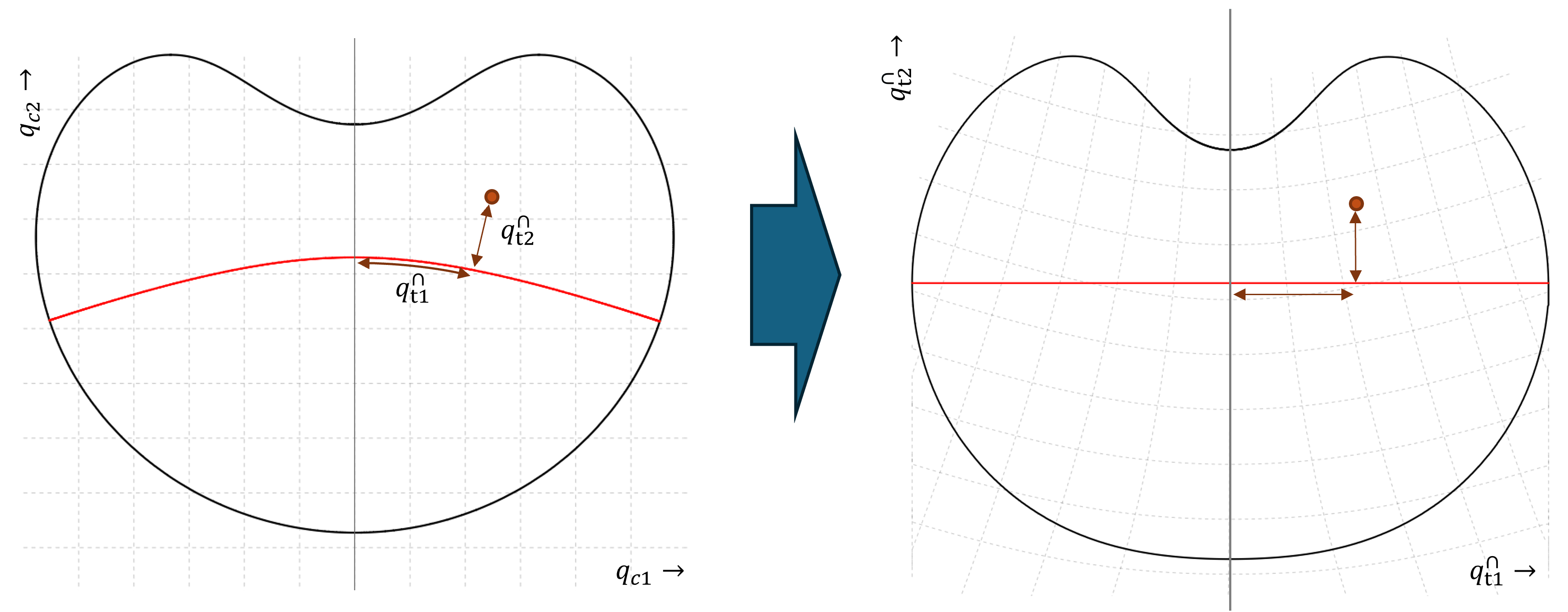}
	\caption{Geometrical representation of the local Frenet-Serret transformation of the configuration space, using the stable "cap" fixed-point trajectory as a continuous reference curve. This framework maps the physical neighborhood into new orthogonal configuration coordinates $(q_{t1}, q_{t2})$ based on arc length and transverse normal displacement.}
	\label{Fig:TrafoFrenet}
\end{figure}

To implement this approach, we parameterize the chosen physical trajectory as a continuous reference curve and define a new set of orthogonal configuration coordinates $(q_{\text{t}1}, q_{\text{t}2})$ for any arbitrary point $(x, y)$ in the Cartesian plane using a local Frenet transformation. 
Within this framework, the coordinate $q_{\text{t}1}$ represents the arc length along the reference trajectory, measured from a designated starting point to the orthogonal projection of the point $(q_{\text{c}1}, q_{\text{c}2})$ onto the curve. 
Complementarily, the second coordinate $q_{\text{t}2}$ is defined as the shortest  distance from the point $(q_{\text{c}1}, q_{\text{c}2})$ to that exact same projection point on the reference curve.
A geometrical representation of this transforation of configuration space is illustrated in Figure \ref{Fig:TrafoFrenet} considering the the cap trajectory as reference.
By using this moving Frenet-Serret frame of the trajectory as a coordinate basis, which consists of the local tangent and normal vectors, one can smoothly map the immediate neighborhood of any curve, and hence the Poincaré sections. 
In this tailored framework, the reference trajectory itself simplifies to the the condition $q_{\text{t}2} = 0$ and consequenlty to the desired formulation as naiv section  that is everywhere strictly parallel to the physical flow of the reference motion.

To complete the canonical transformation for the phase space, the corresponding conjugated momenta $p_{\text{t}1}$ and $p_{\text{t}2}$ must be determined. 
Since the Frenet transformation essentially performs a localized rotation of the unit basis vectors at each point along the reference curve, the new coordinates continue to represent physical distances (arc length and normal displacement).
Consequently, the transformation of the momenta is straightforwardly governed by the associated Jacobian matrix of the coordinate mapping.
 Because the moving frame remains strictly orthogonal, the kinetic energy does not couple the momentum components.
  The momentum $p_{\text{t}2}$, conjugate to the normal distance, directly measures the motion perpendicular to the trajectory, while $p_{\text{t}1}$ tracks the momentum parallel to the progression along the orbit.
With this canonical framework established, the sufficient condition for recording a valid piercing through this trajectory-aligned Poincaré section simplifies once more. 
Since the section is defined by the condition $q_{\text{t}2} = 0$, a directional crossing is uniquely identified whenever the system passes through the reference curve with a strictly positive transverse momentum, $p_{\text{t}2} > 0$ for traditional maps or $p_{\text{t}1} > 0$ for inverted mappings.

\begin{figure}[!bt]
	\centering
	\includegraphics[width=0.6\textwidth]{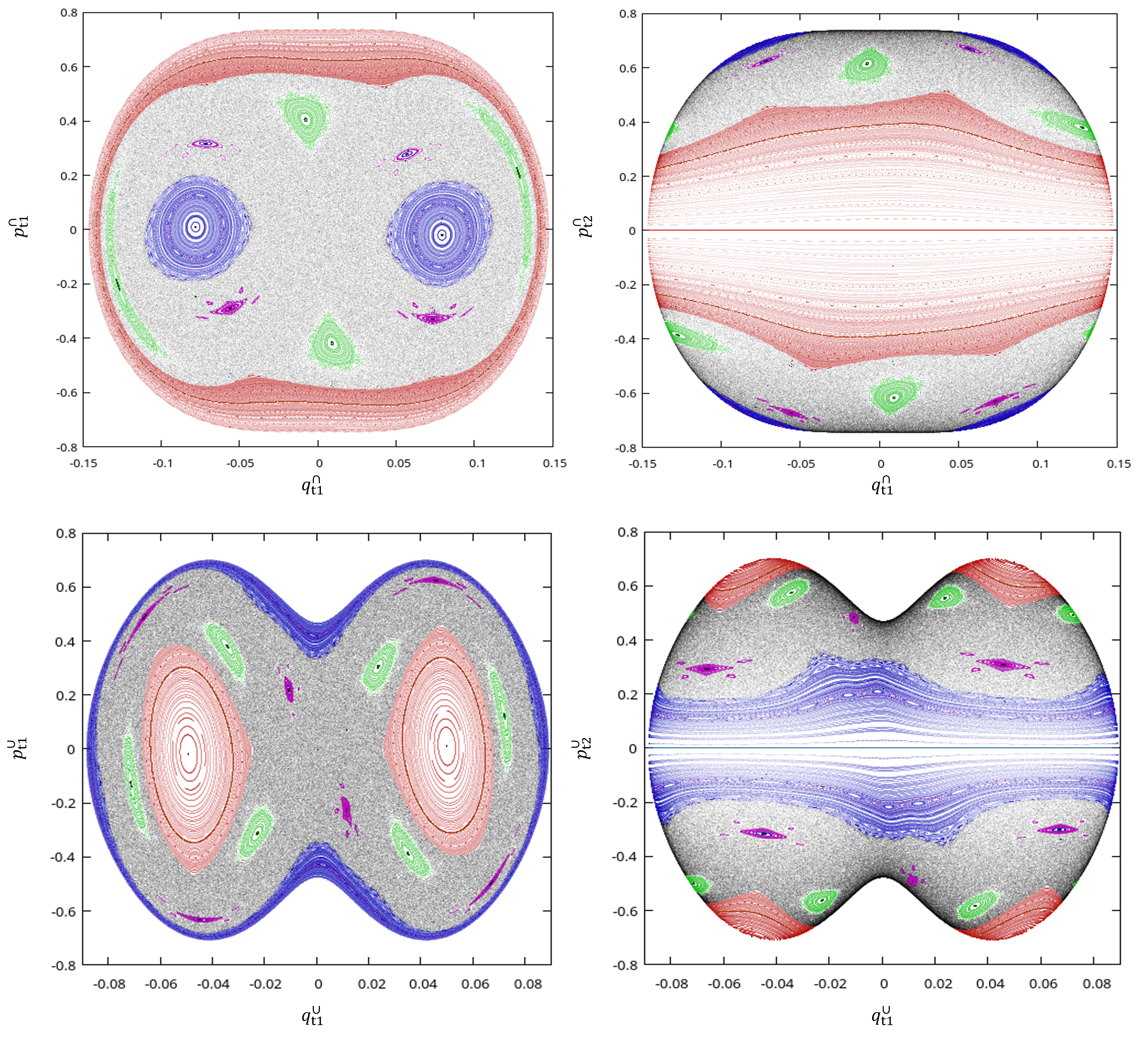}
	\caption{Poincaré mappings utilizing the stable fixed-point trajectories as the sectioning conditions. The top row illustrates the "cap" trajectory section, which features a geometrically locked regular boundary matching the invariant curves of the reference motion. The bottom row depicts the "cup" trajectory section, characterized by a non-convex, indented boundary due to energy conservation rules resulting in two distinct potential wells along the path. The corresponding inverted mappings (right column) unroll the curved invariant manifolds into a flattened, open symmetric layout}
	\label{Fig:PoincareCupCap}
\end{figure}

The obtained mappings for the cup and cap fixed-point trajectories are presented in \fref{Fig:PoincareCupCap}, where again the caotic and different regular trajectories are color coded. 
The corresponding mapping considering the asymmetric trajectory is presented and discussed in detail in the Appendix. 

\subsubsection{cap-trajectory section}
Considering the cap ($\cap$) trajectory as the sectioning condition yields a phase-space representation that exhibits a strong structural similarity to the traditional horizontal section. 
This close alignment, is a direct consequence of the spatial proximity of the two sections in configuration space: the cap trajectory stays geometrically close to the  horizontal section line throughout large portions of the configuration space. Consequently, physical trajectories that pierce the horizontal surface at a steep angle will typically also cross the cap sectioning path under similar conditions. 
Due to the curved nature of the trajectory-aligned section, however, the relative piercing angle between the flow and the coordinate frame dynamically shifts, especially when approaching the outer turnarounds of the configuration space. 
This coordinate distortion manifests visually as a smooth, apparent rotation of the interior island structures when comparing the map obtained from the cap section (\fref{Fig:PoincareCupCap}, top left) to the horizontal map (\fref{Fig:naivPoincareMaps}b). 
The most prominent divergence between the two sections emerges at the boundaries. 
By construction, the cap section completely encapsulates the reference fixed-point trajectory along its entire path. 
As a result, the boundary of the  cap mapping is perfectly delineated by the invariant curves of this stable reference motion itself, wrapping the chaotic sea in a highly regular, geometrically locked frame.

Regarding the inverted mappings, a similar conclusion can be drawn. 
However, due to the displacement of the sectioning path of the horizontal section  to basically the inside of the cup island, the mapping is even more dominated by the associated quasi-periodic islands.
Within this inverted representation, these stable islands no longer appear as isolated localized structures; instead, they manifest themselves as continuous, stratified curves spanning across the entire width of the core region. This topological feature is a direct visual consequence of flattening the curved invariant backbone onto the central coordinate axis ($q_{\text{t}2}=0$), which unrolls the nested regular tracking of the stable manifold into a beautifully symmetric and open layout.

\subsubsection{Cup-trajectory section:}
\label{SubSubSec:CupSections}

Considering the cup ($\cup$) trajectory as the sectioning condition, a fundamentally different mapping is obtained. 
Although the map possesses a clear inversion symmetry due to the spatial symmetry of the sectioning surface, its remaining geometric properties differ substantially from all previous maps.
The most severe difference lies in the non-convex shape of the enveloping boundary curve. 
While all other mappings feature a single global momentum maximum, typically located at the center ($q=0$), the cup section map presents its maximal momentum $p_{\text{t}1}$ twice, near $q_{\text{t}1} \approx \pm 0.043$. 
This striking feature is a direct consequence of energy conservation: the conjugate momentum reaches a maximum where the potential energy along the section path is locally minimized.
While the traditional vertical and horizontal sections as well as the force-line sections all pass directly through the central equilibrium point, this minimum condition is fulfilled only once.
In contrast, because the stable fixed-point trajectories circumvent the equilibrium, their associated sections exhibit a shifted potential minimum away from the coordinate center. 
For the cap ($\cap$) trajectory, the minimal accessible potential remains close enough to $q_{\text{t}1}=0$ that its boundary appears visually regular; however, for the cup section, the spatial separation yields two distinct potential wells, generating the characteristic concave indentation of the boundary.

Comparing the interior of this mapping with the traditional cuts reveals a strong similarity to the vertical section (\fref{Fig:naivPoincareMaps}a). 
As depicted in \fref{Fig:naivSection_Trjectories}d, the cup trajectory extends nearly over the entire vertical range of the accessible configuration space, executing a predominantly vertical movement. 
Consequently, the recorded piercing conditions resemble in large parts those of the classical vertical section. 
A closer look at one side of the map confirms this close relationship: the overall shape of the cap and loop islands matches the structures found in the vertical map. 
Because the cup section tracks this vertical pathway twice (on both the left- and right-hand sides), the total number of primary features is doubled, yielding six loop islands and two central cup islands. 
Less intuitive from a purely visual comparison, however, is the six-fold appearance of the asymmetric island chain.
Comparing the actual paths of the asymmetric fixed-point-one and the cup trajectories in configuration space confirms exactly three intersections per side, thereby explaining the origin of this higher-order resonance structure. 

Again, the enveloping boundary curve of this coordinate representation corresponds precisely to the path of the underlying trajectory itself, causing the stable core structures and primary stability islands associated with this motion naturally to determine the outermost regular boundary of the map. 
This specific geometric ordering provides the ideal foundation for analyzing the corresponding inverted mapping.
Upon considering the inverted mapping, on the other hand,  the invariant fixed-point trajectory is now flattened to form the central horizontal axis ($q_{\text{t}2}=0$), and the quasi-stable island accompaning this orbit are shifted into the focal center of the map.
This representation favours particularily ability to study the internal satellite islands.
As their corresponding trajectories wind around the stable cup-trajectory, the number of geometrical crossing points between stable satelite and the cup trajectory is larger than the number of crossing of the section discussed so far. 

\begin{figure}[!bt]
	\centering
	\includegraphics[width=0.8\textwidth]{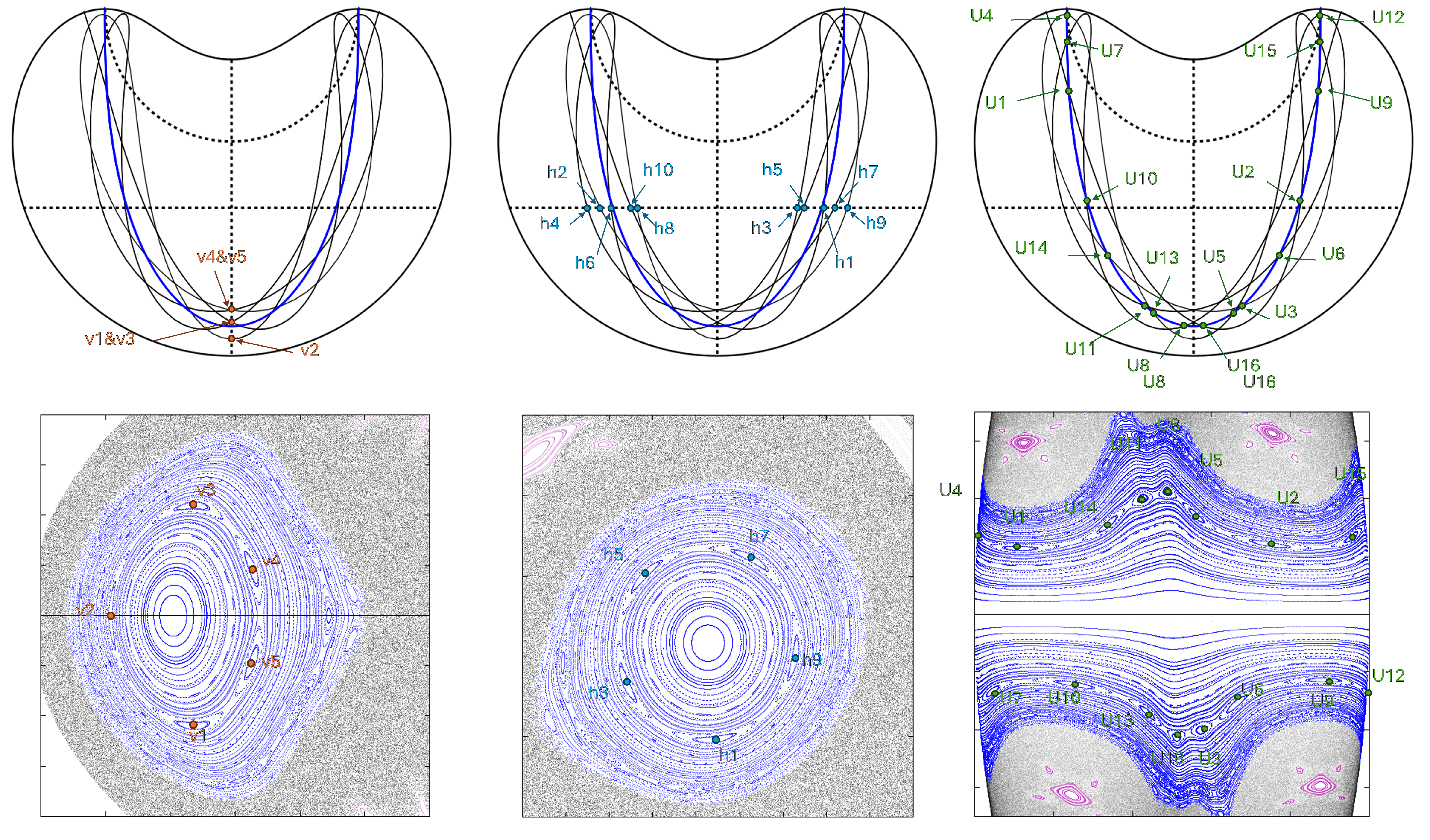}
	\caption{Comparative topological visualization tracking a single satellite trajectory across different Poincaré sections. The trajectory crosses the standard vertical section 5 times $(\Omega_1=5/1)$ and the horizontal section 10 times. By shifting to the curvilinear "cup" section (bottom right), the mapping unlocks 16 distinct crossings $(\Omega_2=16/5)$, explicitly exposing the fine-grained secondary resonances and deep non-linear organization masked in traditional planar cuts. }
	\label{Fig:FigWinding}
\end{figure}

\fref{Fig:FigWinding} demonstrates how a satellite trajectory appears in the vertical, horizontal, and the inverted cup Poincaré mapping. While the trajectory passes the vertical Poincaré section at five distinct points, it pierces the horizontal section ten times, where due to symmetry even piercings occur on one side and odd on the other. On the contrary, the cup section results in 16 piercings.Following the topological classification introduced by Acosta-Zamora et al. \cite{AcostaZamora2024}, a regular phase-space trajectory can be identified by its geometric appearance in a Poincaré section, which is closely related to the inverse of the winding number in knot theory. More specifically, the rational parameter $\Omega = p/q$ is considered, where:
$p$ denotes the total number of distinct clusters, segments, or islands formed by the orbit's intercepts within the Poincaré map and $q$ represents the topological sorting or ``jump'' order of subsequent visits, meaning that the orbit skips $q-1$ segments to reach the next intercept during each iteration.
In our specific case, the vertical as well as the horizontal section yield a global parameter of $\Omega_1 = 5/1$, representing the 5 primary islands or poloidal circuits of the system.
However, this standard view completely masks the internal sub-structures. By shifting to the curvilinear \textit{cup section} the mapping unveils a second-generation hierarchy. 
Here, the trajectory acts as a cable knot with an internal parameter of $\Omega_2 = 16/5$. 
This explicitly demonstrates that the choice of an adapted Poincaré section is crucial: while conventional planar sections only capture the coarse global invariants, the section along the underlaying stability unlocks the fine-grained secondary resonances and the deeper non-linear organization of the phase space.
We evaluated further stable cup trajectories, for which formation of island are visible in the poincaré maps, and found that all of these it holds $\Omega_2=3+\Omega_1^{-1}$. 
Whether or not this observation remains true for other trajectories, in particular the quasiperiodic where $\Omega$ is irrational, or for satelites of the cap or asymetric fix-point trajectories os task of furture works.

In conclusion, utilizing specific invariant fixed-point trajectories to define the Poincaré sectioning condition offers a powerful and highly adaptable diagnostic tool for the  analysis of  multi-degree-of-freedom Hamiltonian systems.
As demonstrated by the cup and cap examples, this approach provides a rigorous framework for unrolling complex, curved invariant manifolds in phase space into a straight, regular coordinate grid. 
Particularly in the context of the inverted mappings, this framework proves invaluable for elucidating the local properties and fine-scale structures of the quasi-periodic trajectories that enclose a stable fixed point. 
By flattening the central reference motion onto a coordinate axis, the local metric distortions are minimized exactly where the nearby regular dynamics reside. 
Consequently, instead of being compressed or sheared near the margins of rigid cuts, the surrounding satellite islands and nested tori are brought directly into the focal center of the mapping. 
This structural unfolding not only yields a highly clear and undistorted representation of the local stability boundaries, but also facilitates a far more detailed study of higher-order resonances, local bifurcations, and the precise tracking of regular orbits in non-linear dynamical systems.

\section{Summary and Outlook}

In this study, we presented a comprehensive comparison of different Poincaré mapping techniques commonly used in the analysis of the planar elastic pendulum. A key observation of our global assessment is that naive, rigid definitions of the sectioning hyperplanes often introduce severe geometric artifacts that can distort the apparent phase-space topology. Crucially, such traditional cuts may completely omit or clip invariant fixed-point trajectories, which should ideally serve as the foundational backbone for any structural representation of the phase space. 

To overcome the limitations of standard tracking, this work introduced—for the first time—a systematic comparison utilizing inverted phase-space mappings. This representation provides a simple yet remarkably efficient methodology to expose and analyze critical dynamical features that are otherwise heavily obscured by the dense packing of invariant curves near the chaotic boundaries of standard sections. By effectively turning the phase space inside out, this inverted approach successfully revealed a hidden topological property: the manifestation of an apparent separatrix trajectory, providing new insights into the global boundary layer organization.

Among the traditional maps investigated, the classical vertical section stood out due to its alignment with the system's inherent physical and geometric symmetries. It uniquely encompasses two fundamental structural features of the underlying field: a continuous fixed-point trajectory and a characteristic line of force. Exploiting these inherent properties, we designed two novel classes of customized Poincaré sections. For each approach, a tailored canonical coordinate transformation was derived, ensuring that the desired sectioning surfaces are elegantly given by simply fixing one associated new coordinate. 

More generally, our global investigation demonstrates that relying on a single slice through the phase space is often insufficient to capture the full complexity of non-linear dynamics. Instead, it is highly recommended to consider at least two distinct Poincaré mappings, ideally chosen such that their underlying sectioning surfaces stand orthogonal to each other within the configuration space. In this regard, coordinate systems derived directly from the physical force-field lines and their orthogonal isopotential counterparts prove to be uniquely suited, as they naturally conform to the invariant metric of the system without introducing external geometric bias.

The first of these novel configurations, the force-line section, successfully resolved the geometric stretching inherent to the physical force field. Although it introduces a coordinate singularity at the potential minimum, it alters the apparent size and distribution of the primary resonance islands to offer a much more balanced view of the phase-space density near the equilibrium, while uncovering a hexagon-like boundary deformation driven by local curvature inflection points. The second configuration, the trajectory-aligned section, demonstrated a unique capability to unroll complex, curved invariant manifolds into a straight, regular grid. By flattening the invariant reference curves onto a central coordinate axis, this approach—especially when combined with the inverted mapping—minimizes metric distortions and brings the delicate surrounding satellite islands and nested tori directly into the focal center of the map, preventing them from being compressed or sheared.

Looking forward, several promising avenues for future research emerge. While the analysis presented in this work focused on a specific, highly illustrative parameter set, extending these new mapping techniques to systematically explore broader parameter scans will be highly advantageous. In particular, utilizing these tailored coordinates to characterize the global order-chaos-order transitions could significantly clarify how resonance islands emerge and submerge under varying non-linear conditions. Furthermore, because the newly defined coordinate systems result in a fundamental deformation of the phase space and its associated isoenergetic subspaces, the intrinsic properties of the trajectories—particularly those characterizing higher-order winding numbers and fractality—should be systematically revisited. Re-evaluating these quantities under the new insights obtained from this multi-mapping comparison promises to uncover a deeper, distortion-free understanding of transport and structural stability in non-linear dynamical systems.




\subsection*{Funding}
This study was financed, in part, by the São Paulo Research Foundation (FAPESP), Brasil. Process Number (2024/21697-4), in part by Conselho Nacional de Desenvolvimento Cientifico e Tecnológico (CNPq), Brasil. Process Number (304662/2025-9)
\printbibliography

@article{CDESOUSA20181110,
title = {Energy distribution in intrinsically coupled systems: The spring pendulum paradigm},
journal = {Physica A: Statistical Mechanics and its Applications},
volume = {509},
pages = {1110-1119},
year = {2018},
issn = {0378-4371},
doi = {https://doi.org/10.1016/j.physa.2018.06.089},
url = {https://www.sciencedirect.com/science/article/pii/S0378437118308148},
author = {M. {C. de Sousa} and F.A. Marcus and I. {L. Caldas} and R. {L. Viana}}
}

@article{Kuznetsov1999,
 author = {Kuznetsov, S. V.},
 title = {The motion of the elastic pendulum.},
 fjournal = {Regular and Chaotic Dynamics},
 journal = {Regul. Chaotic Dyn.},
 issn = {1560-3547},
 volume = {4},
 number = {3},
 pages = {3--12},
 year = {1999},
 language = {English},
 doi = {10.1070/rd1999v004n03ABEH000110},
 zbMATH = {1842521},
 Zbl = {1137.70358}
}

@article{vanderWeele1996,
title = {The order—chaos—order sequence in the spring pendulum},
journal = {Physica A: Statistical Mechanics and its Applications},
volume = {228},
number = {1},
pages = {245-272},
year = {1996},
issn = {0378-4371},
doi = {https://doi.org/10.1016/0378-4371(95)00426-2},
url = {https://www.sciencedirect.com/science/article/pii/0378437195004262},
author = {J.P. {van der Weele} and E. {de Kleine}}
}

@article{AcostaZamora2024,
    author = {Acosta-Zamora, Karla P. and Núñez González, José and González, Ahtziri and Ramos, Eduardo},
    title = {Characterization of a spring pendulum phase-space trajectories},
    journal = {Chaos: An Interdisciplinary Journal of Nonlinear Science},
    volume = {34},
    number = {2},
    pages = {023119},
    year = {2024},
    month = {02},
    issn = {1054-1500},
    doi = {10.1063/5.0183419},
    url = {https://doi.org/10.1063/5.0183419},
    eprint = {https://pubs.aip.org/aip/cha/article-pdf/doi/10.1063/5.0183419/20217589/023119\_1\_5.0183419.pdf},
}

@article{Cuerno1992,
    author = {Cuerno, Rodolfo and {Fernández Rañada}, Antonio and {Ruiz-Lorenzo}, Juan Jesús},
    title = {Deterministic chaos in the elastic pendulum: A simple laboratory for nonlinear dynamics},
    journal = {American Journal of Physics},
    publisher = {American Association of Physics Teachers},
    year = {1992},
    month = jan,
    volume = {60},
    number = {1},
    pages = {73-79},
    note = {7 pages, 6 figures. PACS nrs.: 05.45.+b, 03.20.+i. MR1145312 (92j:70028)},
    issn = {0002-9505},
    url = {https://hdl.handle.net/10016/7159},
    archiveprefix = {e-Archivo}
}

@article{rackauckas2017differentialequations,
  title={Differentialequations.jl--a performant and feature-rich ecosystem for solving differential equations in julia},
  author={Rackauckas, Christopher and Nie, Qing},
  journal={Journal of Open Research Software},
  volume={5},
  number={1},
  pages={15},
  year={2017},
  publisher={Ubiquity Press}
}

@article{CDESOUSA2022128481,
title = {Internal energy exchanges and chaotic dynamics in an intrinsically coupled system},
journal = {Physics Letters A},
volume = {453},
pages = {128481},
year = {2022},
issn = {0375-9601},
doi = {https://doi.org/10.1016/j.physleta.2022.128481},
url = {https://www.sciencedirect.com/science/article/pii/S0375960122005631},
author = {M. {C. de Sousa} and A.B. Schelin and F.A. Marcus and R. {L. Viana} and I. {L. Caldas}}
}

@article{NUNEZYEPEZ1990101,
title = {Onset of chaos in an extensible pendulum},
journal = {Physics Letters A},
volume = {145},
number = {2},
pages = {101-105},
year = {1990},
issn = {0375-9601},
doi = {https://doi.org/10.1016/0375-9601(90)90199-X},
url = {https://www.sciencedirect.com/science/article/pii/037596019090199X},
author = {H.N. Núñez-Yépez and A.L. Salas-Brito and C.A. Vargas and L. Vicente}
}

@misc{Tarigo2026,
      title={Phase-space organization of the elastic pendulum: chaotic fraction, energy exchanges, and the order-chaos-order transition}, 
      author={Juan P. Tarigo and Cecilia Stari and Edson D. Leonel and Arturo C. Marti},
      year={2026},
      eprint={2604.01503},
      archivePrefix={arXiv},
      primaryClass={nlin.CD},
      url={https://arxiv.org/abs/2604.01503}, 
}

@article{vitt1933,
  title={Oscillations of an elastic pendulum as an example of the oscillations of two parametrically coupled linear systems},
  author={Vitt, A. A. and Gorelik, G. S.},
  journal={Zh. Tekh. Fiz.},
  volume={3},
  pages={294},
  year={1933}
}

@article{tselman1970,
  title={On ``pumping-transfer of energy" between nonlinearly coupled oscillators in third-order resonance},
  author={Tsel'man, F. Kh.},
  journal={Journal of Applied Mathematics and Mechanics},
  volume={34},
  number={5},
  pages={916--922},
  year={1970}
}

@article{carretero1994,
  title={Regular and chaotic behaviour in an extensible pendulum},
  author={Carretero-Gonz{\'a}lez, R. and N{\'u}{\~n}ez-Y{\'e}pez, H. N. and Salas-Brito, A. L.},
  journal={European Journal of Physics},
  volume={15},
  number={3},
  pages={139--148},
  year={1994}
}

@article{Anurag2020,
  author  = {Anurag and Basudeb Mondal and Jayanta K. Bhattacharjee and Sagar Chakraborty},
  title   = {Understanding the order-chaos-order transition in the planar elastic pendulum},
  journal = {Physica D: Nonlinear Phenomena},
  volume  = {402},
  pages   = {132256},
  year    = {2020}
}

@article{anurag2022,
  author  = {Anurag and Sagar Chakraborty},
  title   = {Locating order-chaos-order transition in elastic pendulum},
  journal = {Nonlinear Dynamics},
  volume  = {110},
  number  = {1},
  pages   = {37--53},
  year    = {2022}
}

@article{hairer1999stiff, 
  title={Stiff differential equations solved by Radau methods}, 
  author={Hairer, Ernst and Wanner, Gerhard}, 
  journal={Journal of Computational and Applied Mathematics}, 
  volume={111}, 
  number={1-2}, 
  pages={93–111}, 
  year={1999}, 
  publisher={Elsevier}
}

@article{verner2010numerically, 
  title={Numerically optimal Runge–Kutta pairs with interpolants}, 
  author={Verner, James H}, 
  journal={Numerical Algorithms}, 
  volume={53}, 
  number={2-3}, 
  pages={383–396}, 
  year={2010}, 
  publisher={Springer} 
}

@book{arnold1989mathematical,
  title={Mathematical Methods of Classical Mechanics},
  author={Arnold, V.I. and Vogtmann, K. and Weinstein, A.},
  isbn={9781475716948},
  series={Graduate Texts in Mathematics},
  url={https://books.google.com.br/books?id=BF_3sgEACAAJ},
  year={2013},
  publisher={Springer New York}
}

@book{guckenheimer2013nonlinear,
  title={Nonlinear Oscillations, Dynamical Systems, and Bifurcations of Vector Fields},
  author={Guckenheimer, J. and Holmes, P.},
  isbn={9780387908199},
  lccn={97135723},
  series={Applied mathematical sciences},
  year={2013},
  publisher={Springer-Verlag}
}

\clearpage

\appendix
\setcounter{page}{1}
\renewcommand{\thepage}{S\arabic{page}}
\setcounter{section}{0}
\renewcommand{\thesection}{S\arabic{section}}
\setcounter{equation}{0}
\renewcommand{\theequation}{S\arabic{equation}}
\setcounter{figure}{0}
\renewcommand{\thefigure}{S\arabic{figure}}
\setcounter{table}{0}
\renewcommand{\thetable}{S\arabic{table}}


\section{Appendix A: Other Poincaré Sections}

During the preliminary stages of this study, alternative choices of Poincaré sections were explored. Although simpler than the sections presented in the main text, they exhibit several noteworthy topological properties.

\subsection{Polar transformation of the naive vertical section}
\begin{figure*}[bt]
    \centering
    \includegraphics[width=1.0\linewidth]{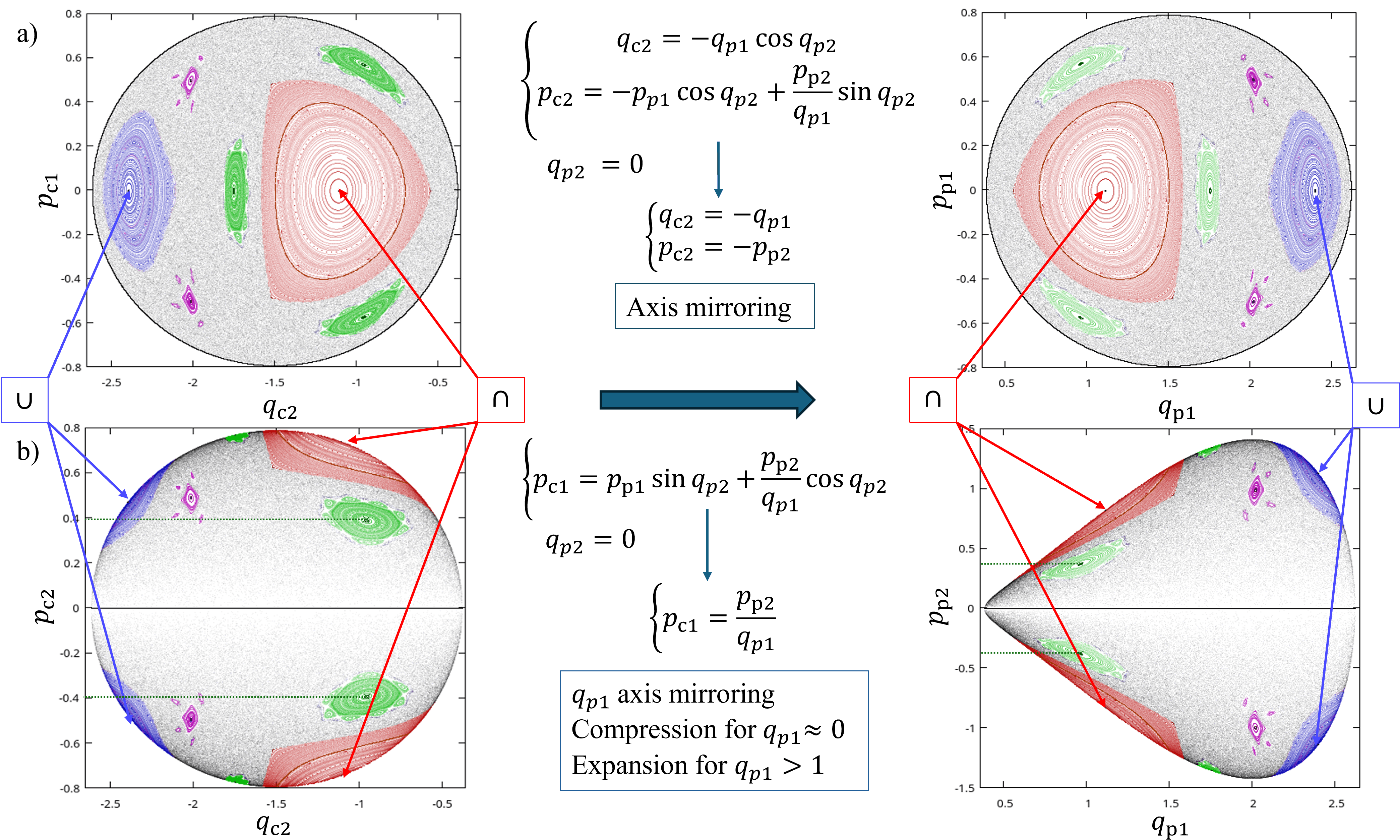}
    \caption{Polar transformation of the Poincaré surface of section at the $q_{\text{c}1} = 0$ cut. The left panels show the phase space structures in the $(q_{\text{c}2}, p_{\text{c}1})$ and $(q_{\text{c}2}, p_{\text{c}2})$ projections. The right panels display the corresponding mapping into polar coordinates $(q_{\text{p}1}, p_{\text{p}1})$ and $(q_{\text{p}1}, p_{\text{p}2})$ evaluated at $q_{\text{p}2} = 0$. The central equations govern the reduction, where panel (a) illustrates the radial momentum mapping dominated by axis mirroring, and panel (b) illustrates the angular momentum mapping showcasing phase-space compression for $q_{\text{p}1} \approx 0$ and expansion for $q_{\text{p}1} > 1$. Regular islands of the "cup" ($\cup$) and "cap" ($\cap$) periodic orbits are tracked using color-matched curves.}
    \label{fig:polar_transform}
\end{figure*}

In polar coordinates, a naive section is obtained by setting $q_{\text{p}2}=0$, which is closely related to the naive Cartesian section $q_{\text{c}1}=0$, albeit with distinct topological particularities. From the relations between Cartesian and polar coordinates, are established by Eqs.~\eqref{Eq:TrafoPolarCard3} giving the corresponding momentum transformations. With $q_{\text{p}2}=0$, these relations become:
\begin{align}
    q_{\text{c}1} &= q_{\text{p}1}\sin(q_{\text{p}2}) \;\rightarrow\; q_{\text{c}1} = 0, \\
    q_{\text{c}2} &= -q_{\text{p}1}\cos(q_{\text{p}2}) \;\rightarrow\; q_{\text{c}2} = -q_{\text{p}1}. \label{qc2polar}
\end{align}
For the momenta:
\begin{align}
    p_{\text{c}1} &= p_{\text{p}1} \sin(q_{\text{p}2}) + \frac{p_{\text{p}2}}{q_{\text{p}1}} \cos(q_{\text{p}2}) \;\rightarrow\; p_{\text{c}1} = \frac{p_{\text{p}2}}{q_{\text{p}1}}, \label{pc1polar}\\
    p_{\text{c}2} &= -p_{\text{p}1} \cos(q_{\text{p}2}) + \frac{p_{\text{p}2}}{q_{\text{p}1}} \sin(q_{\text{p}2}) \;\rightarrow\; p_{\text{c}2} = -p_{\text{p}1}. \label{pc2polar} 
\end{align}
These relations introduce specific geometric modifications. Equation~\eqref{qc2polar} causes a spatial mirroring of the section in $q_{\text{p}2}$ with respect to $q_{\text{c}2}$. For the momentum coordinates, Eq.~\eqref{pc2polar} similarly causes a mirroring of $p_{\text{p}2}$ with respect to $p_{\text{c}2}$; Figure~\ref{fig:polar_transform}a illustrates this mechanism. In Eq.~\eqref{pc1polar}, while there is no mirroring, the presence of the $1/q_{\text{p}1}$ factor induces a spatial compression for values where $q_{\text{p}1} < 1$ and an expansion for $q_{\text{p}1} > 1$. Figure~\ref{fig:polar_transform}b illustrates this process: two islands of the loop trajectory lie at $q_{\text{p}1} = 1$, and their momentum remains $p_{\text{p}1} = p_{\text{c}1} \approx \pm 0.4$.

\begin{figure*}[bt]
    \centering
    \includegraphics[width=0.75\linewidth]{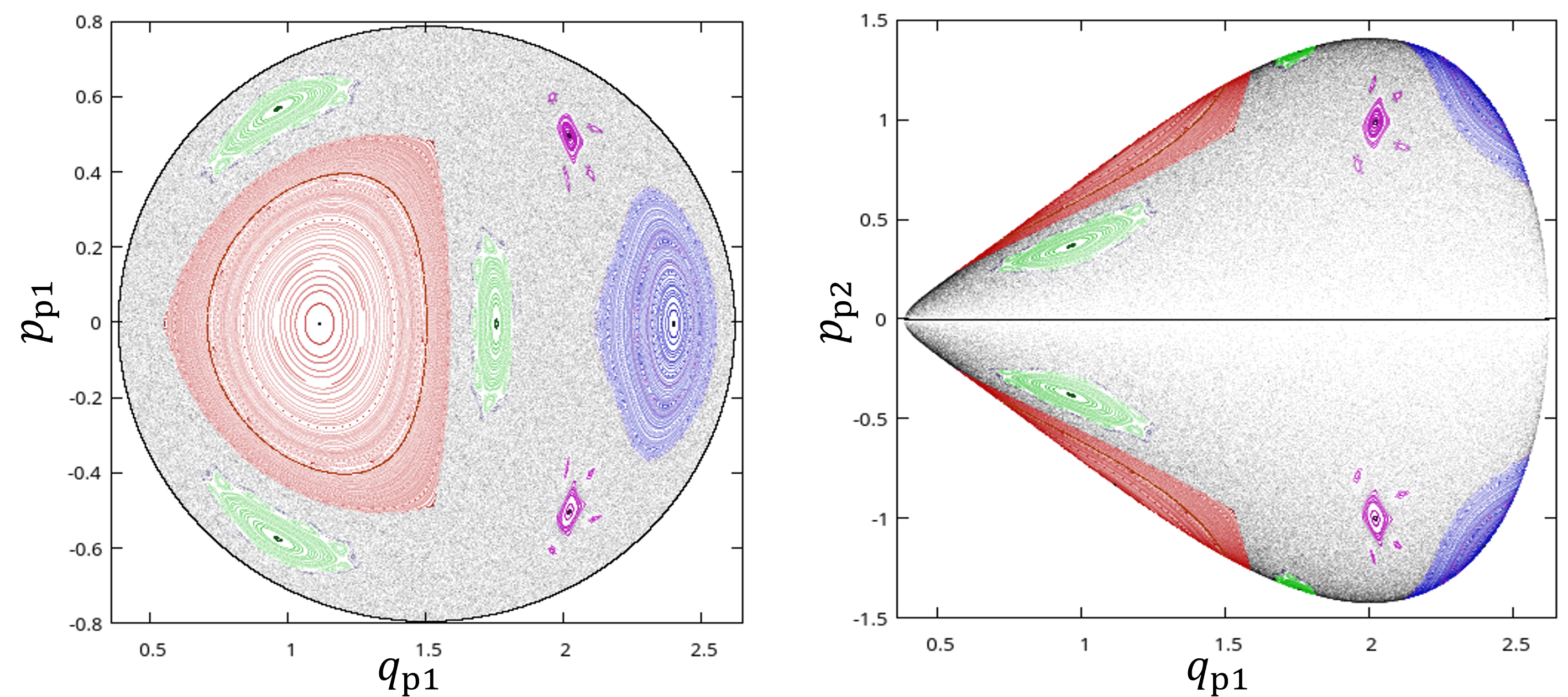}
    \caption{Polar plot of the Poincaré section in the vertical plane at the system's origin.}
    \label{fig:polar_plot}
\end{figure*}

For better visualization, Figure~\ref{fig:polar_plot} presents the naive Poincaré section evaluated at $q_{\text{p}2}=0$.

\subsection{Radial section at the equilibrium length}
\begin{figure*}[bt]
    \centering
    \includegraphics[width=1.0\linewidth]{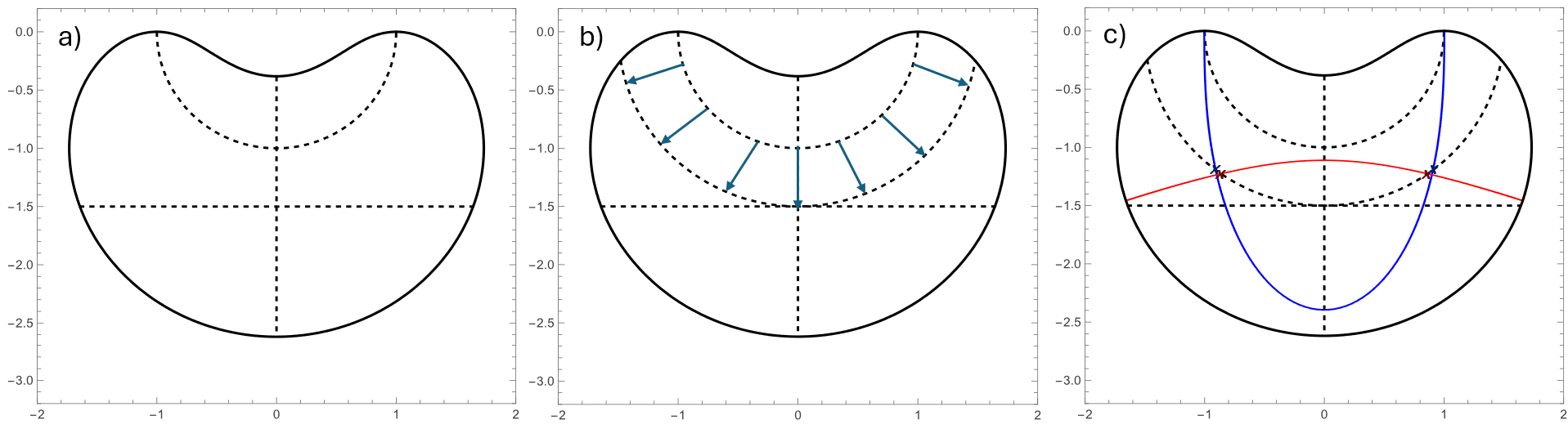}
    \caption{Schematic representation of radial sectioning conditions inside the configuration space boundary. (a) Configuration space slice under the unstretched section condition $q_{\text{p}1}=l_0$. (b) Shift in the sectioning boundary toward the equilibrium length condition $q_{\text{p}1}=l_{\text{eq}}$. (c) Interaction between the cap (red) and cup (blue) fixed-point trajectories and the chosen radial sections, detailing why stability islands shift or change visibility between maps.}
    \label{fig:radial_schematic}
\end{figure*}

Choosing the unstretched natural length $q_{\text{p}1}=l_0$ defines a radial section that cuts the configuration space as shown schematically in Fig.~\ref{fig:radial_schematic}a. A notable limitation of this choice is that the sectioning plane does not intersect several primary orbits; indeed, neither the cap ($\cap$) nor the cup ($\cup$) fixed-point trajectories cross this boundary (Fig.~\ref{fig:radial_schematic}c). To resolve this sampling limitation, we extend the section radius to the equilibrium length $q_{\text{p}1}=l_{\text{eq}}$ (Fig.~\ref{fig:radial_schematic}b). Remarkably, this radial expansion not only uncovers the fixed-point trajectories within the phase-space domain (whose geometric interactions are tracked in Fig.~\ref{fig:radial_schematic}c), but it also generates an apparent separatrix similar to that found in the horizontal section. The physical origin of this apparent separatrix is localized to regions where trajectories evolve nearly parallel to the sectioning arc, particularly near the coordinate point $(q_{\text{c}1}, q_{\text{c}2}) = (0, l_{\text{eq}})$. This topological feature is completely absent in the initial $l_0$ radial representation.

\begin{figure*}[bt]
    \centering
    \includegraphics[width=0.75\linewidth]{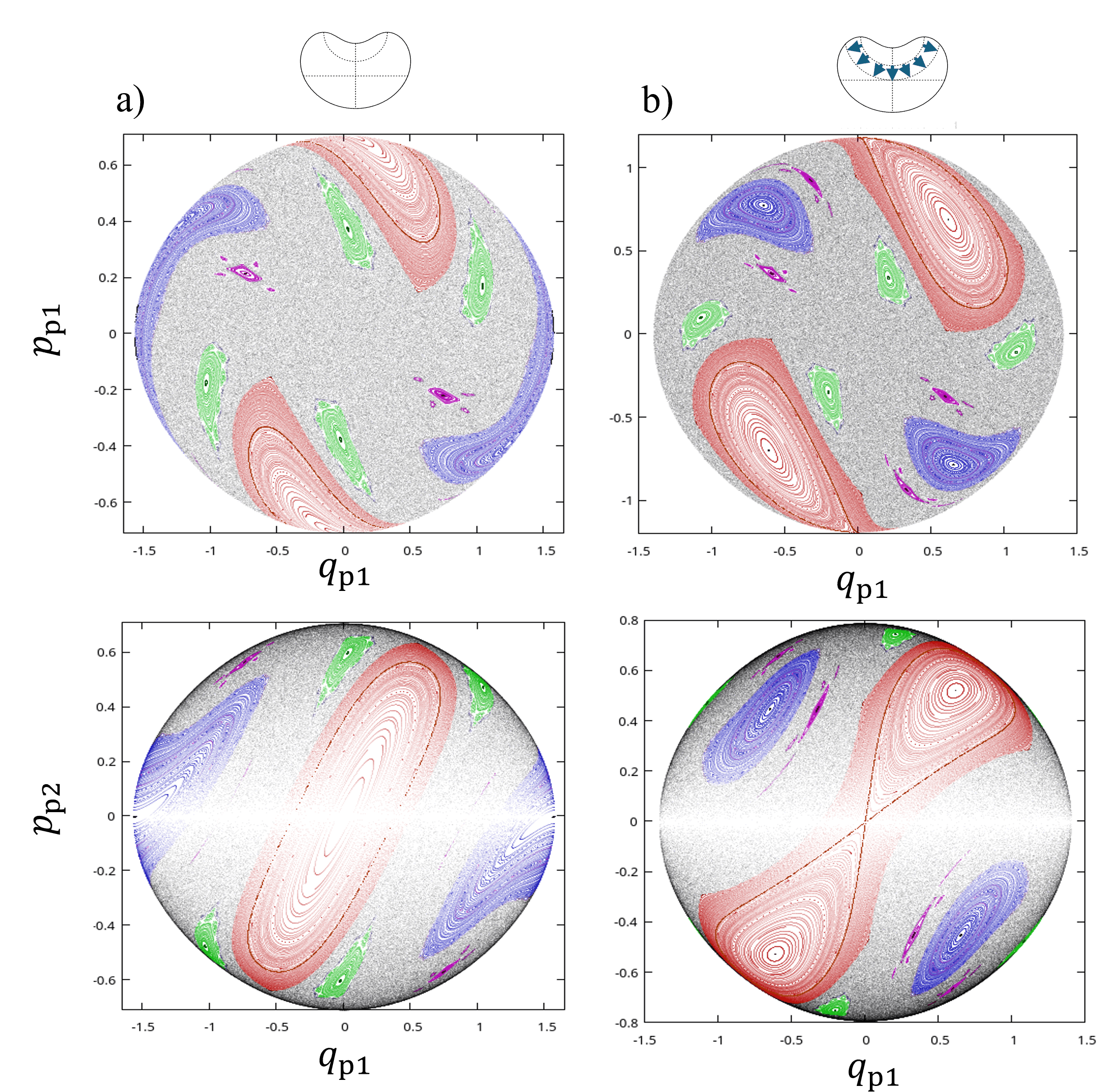}
    \caption{Comparison between the radial Poincaré surfaces of section evaluated at (a) the unstretched natural length $q_{\text{p}1}=l_0$ and (b) the static equilibrium length $q_{\text{p}1}=l_{\text{eq}}$. For both cases, the top row displays the traditional $(q_{\text{p}2}, p_{\text{p}1})$ phase-space projection, while the bottom row shows the corresponding inverted $(q_{\text{p}2}, p_{\text{p}2})$ mapping.}
    \label{fig:radial_eq}
\end{figure*}

The radial expansion introduces significant topological modifications in the projected phase space. While the global arrangement of the stability islands remains qualitatively similar, the expansion drives a noticeable migration of these structures. For the cap ($\cap$) family, the underlying fixed-point trajectory enters the phase space, causing its periodic island to undergo an apparent bifurcation at $q_{\text{p}2}=0$, giving rise to the apparent separatrix. Concurrently, the cup ($\cup$) stability island fully penetrates the accessible phase space, marking the appearance of the central cup fixed-point trajectory. Interestingly, the loop and asymmetric islands exhibit mutually inverted behaviors under this transformation. The loop structure displays four islands in the $q_{\text{p}1}=l_0$ section, which reduce to two islands in the $q_{\text{p}1}=l_{\text{eq}}$ cut. In contrast, the asymmetric family transitions from two visible islands in the initial section to a configuration of four distinct islands in the equilibrium section; this behavior is captured in the traditional $(q_{\text{p}2}, p_{\text{p}1})$ projections shown in the top rows of Figs.~\ref{fig:radial_eq}a and \ref{fig:radial_eq}b. In the corresponding inverted maps (which display the $(q_{\text{p}2}, p_{\text{p}2})$ projections shown in the bottom rows of Figs.~\ref{fig:radial_eq}a and \ref{fig:radial_eq}b), the primary structural distinction is that the cap ($\cap$) and cup ($\cup$) stability islands become fully integrated into the projected phase-space domain.

\subsection{Force-line sections for other $q_{\text{f1}}$ values}

\begin{figure*}[bt]
    \centering
    \includegraphics[width=1.0\linewidth]{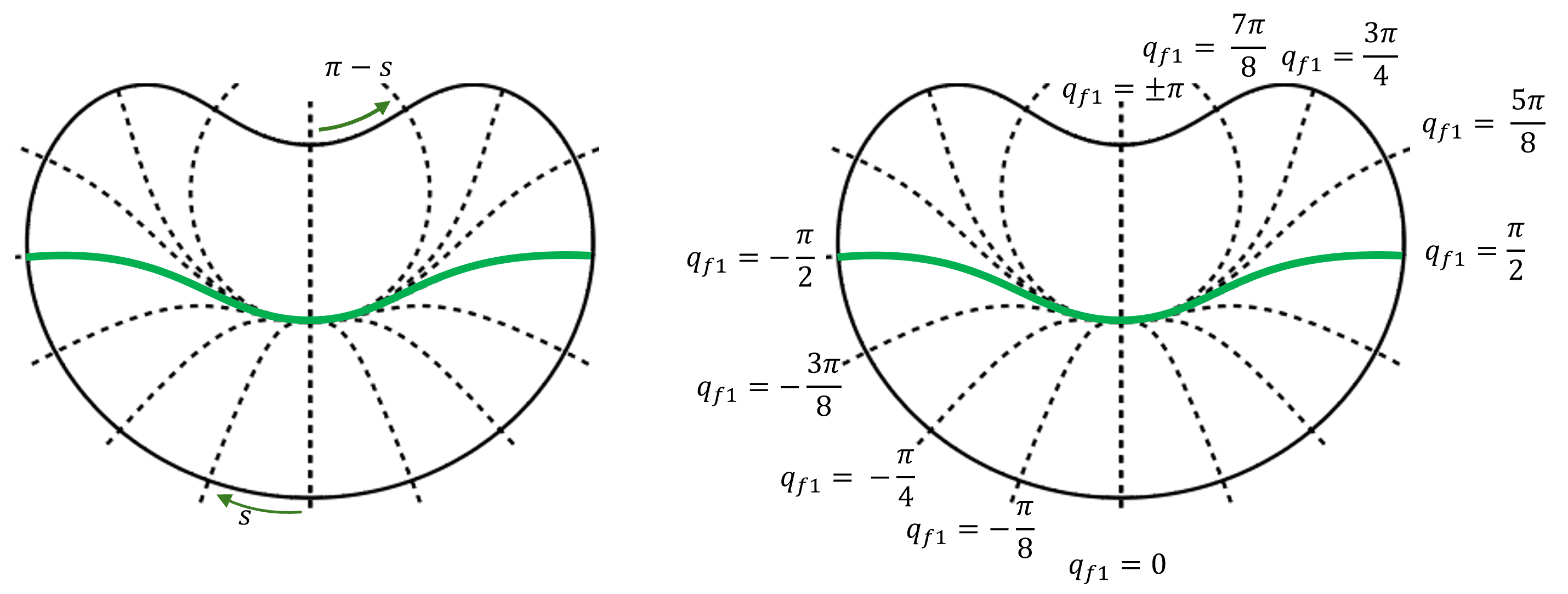}
    \caption{Spatial representation of the rotational sweep of the force-line coordinate $q_{f1}$ across the normalized interval $[0, \pm\pi]$ within the configuration space. }
    \label{Fig:SI:qf1rotation}
\end{figure*}

Following the methodology established in Section \ref{SubSec:ForceSections}, the analysis is extended to other values of $q_{\text{f1}}$ by sweeping the coordinate across the interval $q_{\text{f1}} \in [0, \pm\pi]$. Due to the spatial symmetry of the problem, the left arm rotates clockwise from $0$ to $-\pi/2$, and the right arm rotates from $\pm\pi$ to $\pi/2$, as illustrated in Figure~\ref{Fig:SI:qf1rotation}. Crucially, the boundaries of this rotational sweep correspond to distorted analogs of traditional, naive sectioning hyperplanes. The symmetric vertical pairing $q_{\text{f1}} \in [0, \pm\pi]$ closely mirrors a naive polar section at $q_{\text{p}2} = 0$. Conversely, the orthogonal pairing $q_{\text{f1}} \in [-\pi/2, \pi/2]$ acts as a highly distorted version of the naive Cartesian section at $q_{\text{c}2} = -1-\omega^2$. However, in contrast to the Cartesian cut—which inherently truncates trajectories that do not cross the equilibrium length—the force-line section bends alongside the system's natural energy boundaries, ensuring that all accessible trajectories are mapped entirely on the Poincaré plane.

\begin{figure*}[bt]
    \centering
    \includegraphics[width=1.0\linewidth]{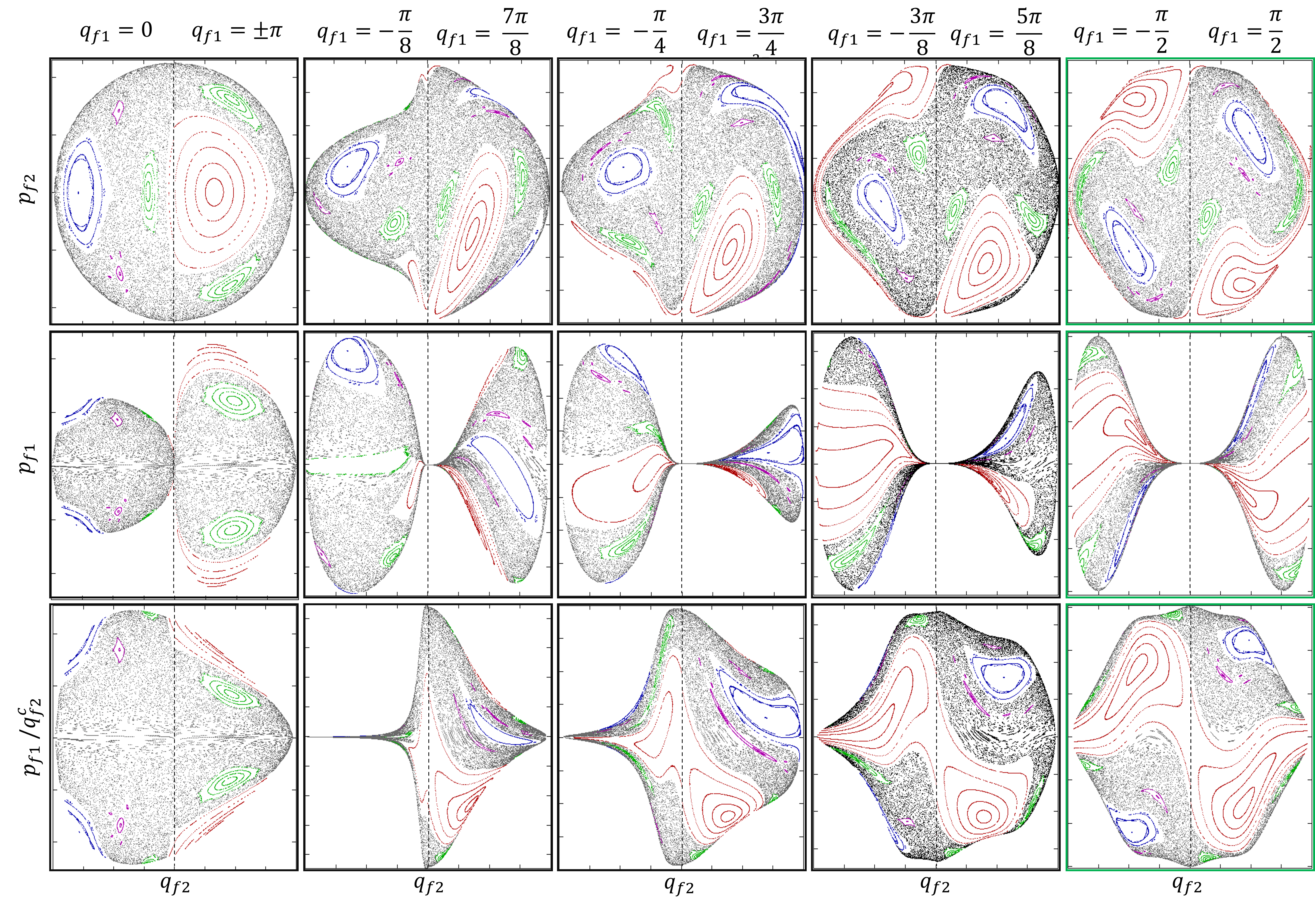}
    \caption{Progression of Poincaré section maps constructed via the continuous rotational sweep of the force-line coordinate $q_{f1}$. The sequence tracks the structural migration, bifurcation, and eventual symmetric re-anchoring of the primary stability islands and chaotic sea as the section transitions between its vertical and orthogonal analogs. }
    \label{FIG:SI:poincare_sec_Snu}
\end{figure*}

Figure~\ref{FIG:SI:poincare_sec_Snu} presents the Poincaré sections constructed via this rotation of $q_{\text{f1}}$. Initially, in the $q_{\text{f1}} = 0$ cut, the cap orbit manifests as a massive island centered on the equator ($p_{\text{f1}}=0$). As the section rotates toward $\pm\pi/2$, this central mass splits and migrates outward, eventually anchoring at the absolute momentum extremes in the final configuration. Conversely, the loop islands originate around the cap island in the $q_{\text{f1}} = 0$ cut. As the sectioning surface rotates, they migrate steadily inward. By the time the section reaches $q_{\text{f1}} = \pm\pi/2$, the green loop islands have settled near the equatorial plane. Furthermore, the cup trajectory island starts on the equator of the $q_{\text{f1}} = 0$ half-section, and it remains broadly bound to the equatorial plane throughout the rotation. It distorts through the asymmetric intermediate steps before mirroring symmetrically on both the left and right halves of the final Cartesian analog. Finally, the asymmetric trajectories (purple), residing in the intermediate transitional zones alongside the cup orbits, are sliced at continuously changing oblique angles. This rotational shear physically fractures their invariant tubes into multiple distinct satellite clusters during the asymmetric intermediate steps before they settle into symmetric pairs in the final configuration.

In the inverted force-line map, utilizing the alternative conjugate momentum successfully unfolds the central singularity into a continuous lobed structure, thereby clarifying the lateral trajectory routing. Starting as the dominant upper and lower volume of the right-hand lobe in the $q_{\text{f1}} = 0$ cut, the cap trajectories migrate to occupy the vast central bulk of both the left and right lobes in the symmetric final configuration. Conversely, the loop islands are initially located at the central area of the right lobe. As the sectioning surface rotates, the rotational shear forces the green islands to migrate outward along the boundary, ultimately driving them to the extreme lateral tips of the bounding envelope in the final orthogonal section. The blue cup trajectory starts at the lateral edge of the left lobe and migrates vertically during the rotation, settling at the top and bottom edges of the left and right lobes in the final configuration. Similarly, the purple asymmetric trajectories migrate through the intermediate chaotic regions; unhindered by any central spatial collapse, their topological splitting becomes distinctly visible as they fracture across the lobed structure during the rotational shear, ultimately settling symmetrically near the cup islands in the final orthogonal cut.

In the normalized map ($1/q_{\text{f2}}^3$), the spatial blow-up transforms the geometry into a four-lobed, cross-like topology for the chaotic sea, where the rotational sweep explicitly visualizes how the different trajectories are routed into specific spatial quadrants. Dominating the massive right bulk in the initial $q_{\text{f1}} = 0$ cut, the rotational sweep mathematically routes the cap islands vertically, ultimately isolating them strictly within the top and bottom vertical lobes in the final configuration. Conversely, originating confined in the far-left tail, the blue cup trajectories are routed horizontally as the section rotates to $\pm\pi/2$, eventually settling into the broad left and right horizontal lobes. By the end of the sweep, the green loop and purple asymmetric trajectories are caught within the extreme topological shear zones. As the section rotates, the green loop islands are pushed to the absolute extreme lateral tips of the left and right horizontal lobes, while the purple asymmetric trajectories fragment and compress within the diagonal separatrix boundaries rather than occupying the bulk volume of the quadrants.

\subsection{Frenet-Serret Section: Asymmetric Reference Path}

Another notable period-one orbit is the asymmetric ($\upsilon$) trajectory, which can be physically interpreted as a deformed cup trajectory. By adopting the approach from Section \ref{SubSec:TrajSections} and employing this asymmetric trajectory as the sectioning condition, we observe morphological similarities between the asymmetric and cup sections that stem directly from this underlying geometric relationship.

The primary similarity lies in the non-convex shape of the enveloping boundary curve. Specifically, the asymmetric section map exhibits its maximal momentum $p_{\text{t}1}$ twice, near $q_{\text{t}1} \approx \pm 0.055$. As detailed in Section \ref{SubSubSec:CupSections}, this specific boundary profile is a direct consequence of energy conservation along the reference fixed-point trajectory. Furthermore, similar to the cap trajectory, the asymmetric orbit spans nearly the entire vertical range of the accessible configuration space, indicating predominantly vertical motion. Consequently, the resulting piercing conditions share kinematic characteristics with the vertical section, albeit with a complete absence of spatial symmetry in the resulting section map.

Regarding the phase-space topology, the cap and loop islands retain their previously observed multiplicity. In contrast, the cup island now manifests three times within the asymmetric section, reinforcing the intrinsic connection between the cup period-one orbit and the asymmetric trajectory. Although the cap and loop islands are conserved in number compared to the vertical or cup sections, their shapes are noticeably deformed. This distortion is a direct consequence of the symmetry breaking inherent in the asymmetric reference path.

Finally, the enveloping boundary curve in this coordinate representation corresponds precisely to the physical path of the underlying reference trajectory. When projected into the inverted map, this boundary becomes wrapped in the central region, creating a spatial focus. Because of this localized focusing, the satellite islands associated with that specific fixed-point trajectory appear more frequently than the satellites of its chiral-opposite counterpart.
\begin{figure*}[bt]
    \centering
    \includegraphics[width=1.0\linewidth]{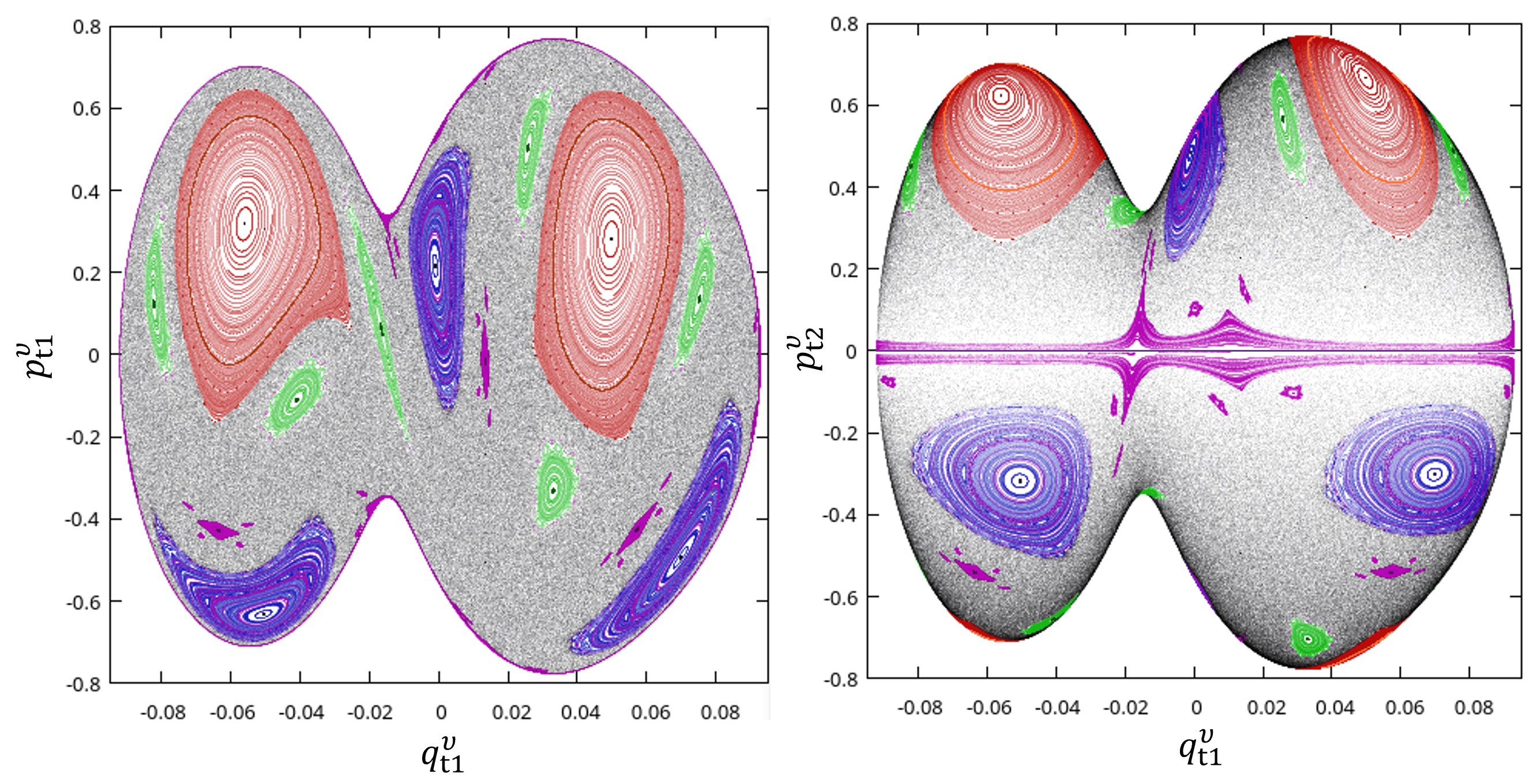}
    \caption{Poincaré section and corresponding inverted mapping constructed utilizing the asymmetric fixed-point-one trajectory as the reference Frenet-Serret path. The maps highlight the deformed secondary island structures and the central spatial focusing caused by the localized wrapping of the asymmetrical reference boundary.}
    \label{FIG:asymm_map}
\end{figure*}
\end{document}